\documentclass[a4paper,11pt]{article}

\usepackage{jheppub} 
\usepackage{graphicx}
\usepackage{booktabs}
\usepackage{multirow}
\usepackage{subfig}
\usepackage{float}
\usepackage{listings}
\usepackage{placeins}
\usepackage{morefloats}
\usepackage{lineno}
\usepackage{color}

\definecolor{dkgreen}{rgb}{0,0.6,0}
\definecolor{gray}{rgb}{0.5,0.5,0.5}
\definecolor{mauve}{rgb}{0.58,0,0.82}

\title{A determination of $m_c(m_c)$ from HERA data using a matched heavy-flavor scheme}

\author[]{xFitter Developers' team:}
\author[a,1]{Valerio Bertone,\note{Corresponding author.}}
\author[b]{Stefano Camarda,}
\author[a]{Amanda Cooper-Sarkar,}
\author[c]{Alexandre Glazov,}
\author[d]{Agnieszka {\L}uszczak,}
\author[c]{Hayk Pirumov,}
\author[c]{Ringaile Pla\v{c}akyt\.{e},}
\author[e]{Klaus Rabbertz,}
\author[a]{Voica Radescu,}
\author[a]{Juan Rojo,}
\author[f]{Andrey Sapranov,}
\author[c]{Oleksandr Zenaiev,}
\author[c]{and Achim Geiser.}

\affiliation[a]{Rudolf Peierls Centre for Theoretical Physics,
1 Keble Road, University of Oxford,
OX1 3NP Oxford, UK}
\affiliation[b]{European Organization for Nuclear Research (CERN), CH-1211 Geneva 23, Switzerland}
\affiliation[c]{Deutsches Elektronen-Synchrotron (DESY), Notkestrasse
  85, D-22607 Hamburg, Germany}
\affiliation[d]{T. Kosciuszko Cracow University of Technology, Institute of Physics, st. Podchorozych 1, 30-084 Cracow}
\affiliation[e]{Karlsruher Institut f\"ur Technologie (KIT), Karlsruhe, Germany}
\affiliation[f]{Joint Institute for Nuclear Research (JINR),
  Joliot-Curie 6, 141980, Dubna, Moscow Region, Russia}

\emailAdd{valerio.bertone@physics.ox.ac.uk}

\abstract{The charm quark mass is one of the fundamental parameters of
  the Standard Model Lagrangian. In this work we present a
  determination of the $\overline{\rm MS}$ charm mass from a fit to
  the inclusive and charm HERA deep-inelastic structure function
  data. The analysis is performed within the {\tt xFitter} framework,
  with structure functions computed in the FONLL general-mass scheme
  as implemented in {\tt APFEL}.  In the case of the FONLL-C scheme,
  we obtain
  $ m_c(m_c)=1.335 \pm
  0.043\mbox{(exp)}^{+0.019}_{-0.000}\mbox{(param)}^{+0.011}_{-0.008}\mbox{(mod)}^{+0.033}_{-0.008}\mbox{(th)}
  \mbox{ GeV.}  $
  We also perform an analogous determination in the
  fixed-flavor-number scheme at next-to-leading order, finding
  $ m_c(m_c) =
  1.318\pm0.054\mbox{(exp)}^{+0.011}_{-0.010}\mbox{(param)}^{+0.015}_{-0.019}\mbox{(mod)}^{+0.045}_{-0.004}\mbox{(th)}\mbox{
    GeV,} $
  compatible with the FONLL-C value. Our results are consistent with
  previous determinations from DIS data as well as with the PDG world
  average.}

\begin{document}
\vspace{-1.0cm}
\begin{flushright}
OUTP-16-10P \\
DESY Report-16-078
\end{flushright}
\begin{figure}[h]
\includegraphics[width=.22\textwidth]{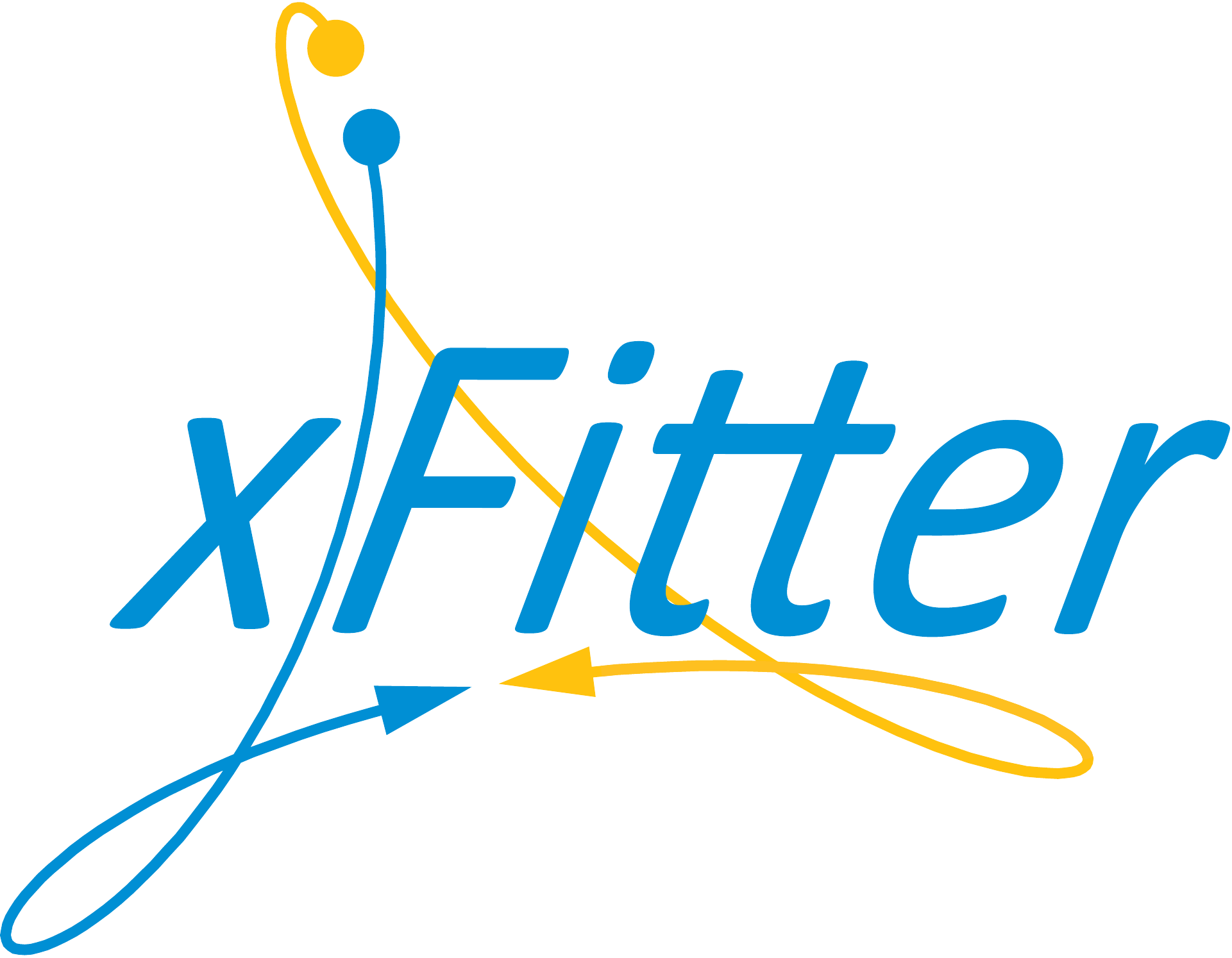}\\

\vspace{-0.5cm}
\includegraphics[width=.22\textwidth]{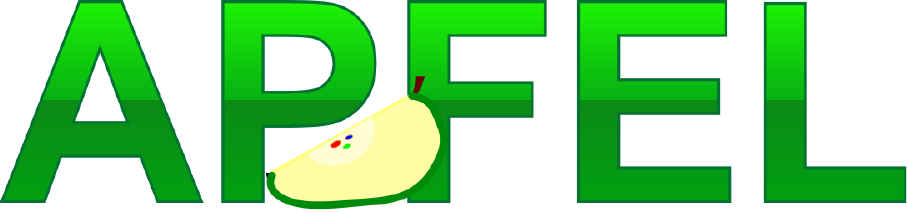}
\end{figure}

\maketitle

\flushbottom


\section{Introduction}\label{sec:intro}

The masses of the heavy quarks, charm, bottom and top, are fundamental
parameters of the Standard Model~\cite{Moch:2014tta}.
A precise determination of their values is of utmost importance; as an
example, the fate of the electroweak vacuum depends crucially on the
exact value of $m_t$~\cite{Degrassi:2012ry}.
In the case of the charm quark, since its mass $m_c$ is larger than
the scale $\Lambda_{\rm QCD}$ of Quantum Chromodynamics (QCD), its
value is a direct input of many perturbative calculations involving
charm quarks in the initial and/or in the final state.

Differences in the value of the charm quark mass and in the
treatment of its effects in deep-inelastic-scattering structure functions can
lead to differences in modern analyses of parton
distribution functions
(PDFs)~\cite{Ball:2014uwa,Harland-Lang:2014zoa,Dulat:2015mca,Alekhin:2013nda,Abramowicz:2015mha},
with implications for precision phenomenology at the Large
Hadron Collider (LHC).
As a consequence, a high-precision determination of the charm quark
mass is of interest both in principle, as a fundamental test of the
Standard Model and a measurement of one of its fundamental parameters,
and in practice, as input for LHC calculations.

The current global-average value of the charm mass in the
$\overline{\rm MS}$ renormalization scheme is
$m_c(\mu_R = m_c)=1.275 \pm 0.025$ GeV~\cite{Agashe:2014kda}, where
the result is dominated by high-precision data from charm production
in $e^+e^-$ collisions.
It is therefore interesting to provide alternative determinations of
the charm mass from other processes, both to test the robustness of
the global average and to attempt to further reduce the present
uncertainty.

A process directly sensitive to the charm mass is open-charm
production in lepton-proton deep-inelastic scattering (DIS).
This process has been measured with high accuracy at the HERA collider and the
results of different measurements implying various charm-tagging
techniques are combined~\cite{Abramowicz:1900rp}. The charm
contribution to the inclusive structure functions can be determined
through the measurement of the charm-pair production cross section.
In addition, the final combination of inclusive measurements from Runs
I and II at HERA has been recently presented
in~\cite{Abramowicz:2015mha}.

DIS structure functions can be described using a variety of
theoretical schemes, including the fixed-flavor number (FFN) scheme,
where charm mass effects are included to a fixed perturbative order,
the zero-mass variable-flavor number (ZM-VFN) scheme that neglects
power-suppressed terms in the charm mass but resums to all orders
large collinear logarithms, and the so-called matched general-mass
variable-flavor-number (GM-VFN) schemes, which interpolate smoothly
between the two regimes.  A recent discussion and summary of the
application of these schemes to heavy-flavor data at HERA can be
found $e.g.$ in \cite{Behnke:2015qja}.

Examples of matched general-mass schemes in electro-, photo- and hadroproduction
include FONLL~\cite{Forte:2010ta,Ball:2013gsa,Cacciari:1998it},
TR~\cite{thornehq,Thorne:,Thorne:2014toa}, ACOT~\cite{Guzzi:2011ew},
and a scheme generically referred to as
GMVFNS~\cite{Kramer:2003jw,Kniehl:2009mh,Kneesch:2007ey,Kniehl:2004fy,Kniehl:2012ti}.
In this work we will mostly concentrate on the FONLL scheme and on
its implications for the determination of the charm mass. For
the sake of comparison with previous studies~\cite{Alekhin:2012un,Abramowicz:1900rp,Alekhin:2012vu,Alekhin:2013qqa}, a determination
of the charm mass in the FFN scheme at NLO is also performed.

The original formulation of the FONLL  general-mass scheme for DIS
structure functions was derived in the pole (on-shell)
heavy quark scheme~\cite{Forte:2010ta}.
In Ref.~\cite{Alekhin:2010sv} it was shown how DIS structure functions
in the FFN scheme can be modified to include $\overline{\rm MS}$
heavy-quark masses.
The same scheme conversion can be applied to any GM-VFN scheme,
and in this  work we provide  the relevant expressions for FONLL structure
functions with $\overline{\rm MS}$ running masses.
The main advantage of the use of  $\overline{\rm MS}$ masses
is the possibility of direct
connection with the precise determinations from low-energy
experimental data~\cite{Agashe:2014kda}.

In this work we will use the {\tt xFitter} open-source
framework~\cite{Alekhin:2014irh} (previously known as {\tt HERAfitter})
  to extract the
$\overline{\rm MS}$ charm mass from a PDF fit to the most up-to-date
inclusive and charm data from HERA.
Structure functions are computed using the FONLL scheme as implemented
in the {\tt APFEL}~\cite{Bertone:2013vaa} code.
Our results have been obtained employing the most accurate
perturbative calculations presently available and will include a
detailed characterization of the different sources of uncertainties on
$m_c(m_c)$ from data, theory and fitting methodology.
As we will show, the results are consistent with the global PDG average as
well as with previous determinations based on the
FFN~\cite{Alekhin:2012un,Abramowicz:1900rp,Alekhin:2012vu,Alekhin:2013qqa}
and in the S-ACOT~\cite{Gao:2013wwa} schemes.\footnote{See also~\cite{Harland-Lang:2015qea} for a recent determination of the pole charm mass from a global
PDF fit.}
The uncertainty in our results turns out to be competitive with that
of previous determinations based on DIS structure functions.

The outline of this paper is the following.
In Sect.~\ref{sec:fonll} we discuss how FONLL can be formulated in
terms of $\overline{\rm MS}$ masses and present a benchmark of its
implementation in {\tt APFEL}.
In Sect.~\ref{sec:fitsettings} we describe the settings of the PDF
fits and the treatment of the uncertainties.
Results for the determination of $m_c(m_c)$ are presented in
Sect.~\ref{sec:results}, where we also compare with previous
determinations.
We conclude and discuss possible next steps in
Sect.~\ref{sec:summary}.

\section{FONLL with $\overline{\rm MS}$ heavy-quark masses}\label{sec:fonll}

In this section we discuss how the FONLL general-mass
variable-flavor-number scheme for DIS structure functions can be
expressed in terms of $\overline{\rm MS}$ heavy-quark masses. We also
describe the subsequent implementation in the public code {\tt APFEL},
and present a number of benchmark comparisons with other public codes.

In general, higher-order calculations are affected by ambiguities in
the prediction for the physical quantities due to the choice of the
subtraction scheme used to remove divergences. In fact, different
prescriptions imply different numerical values of the parameters of
the underlying theory.

As far as the mass parameters are concerned, the pole mass definition
is usually more common in the calculation of massive higher-order QCD
corrections to heavy-quark production processes. The main reason for
this is that the pole mass is, by its own definition, more closely
connected to what is measured in the experiments. On the other hand,
it is well known that observables expressed in terms of the pole mass
present a slow perturbative convergence. This is caused by the fact
that the pole mass definition suffers from non-perturbative effects
which result in an intrinsic uncertainty of order
$\Lambda_{\rm QCD}$~\cite{Chetyrkin:1999qi}. The $\overline{\rm MS}$
scheme, which stands for modified minimal subtraction scheme, is
instead free of such ambiguities and as a matter of fact massive
computations expressed in terms of heavy-quark masses normalized in
this scheme present a better perturbative
convergence~\cite{Alekhin:2010sv}. As a consequence, the results
obtained in the $\overline{\rm MS}$ scheme are more appropriate to
achieve a reliable determination of the numerical value of the charm
mass.

The FONLL scheme, as any other GM-VFN scheme, aims at improving the
accuracy of fixed-order calculations at high scales by matching them
to resummed computations. In DIS this results in the combination of
massive (fixed-order) calculations, that are more reliable at scales
closer to the heavy-quark masses, with resummed calculations that are
instead more accurate at scales much larger than the heavy-quark
masses. However, in the original derivation, the massive component of
the FONLL scheme was expressed in terms of the pole
masses~\cite{Forte:2010ta}.

It is then one of the goals of this paper to provide a full
formulation of the FONLL scheme applied to DIS structure functions in
terms of $\overline{\rm MS}$ masses. A detailed discussion on such a
formulation is given below in Sect.~\ref{sec:implementation}. Here, we
limit ourselves to describing the main steps needed.

The generic form of the DIS structure functions in the FONLL approach
applied to charm production is:
\begin{equation}\label{eq:FONLLdef}
\begin{array}{rcl}
F(x,Q,m_c) &=& F^{(3)}(x,Q,m_c) + F^{(d)}(x,Q,m_c)\\
\\
F^{(d)}(x,Q,m_c) &=& F^{(4)}(x,Q) - F^{(3,0)}(x,Q,m_c)\,,
\end{array}
\end{equation}
where $x$, $Q$, and $m_c$ are the Bjorken variable, the virtuality of
the photon, and the mass of the charm quark, respectively. In
eq.~(\ref{eq:FONLLdef}) the three-flavor structure function $F^{(3)}$
is evaluated retaining the full charm-mass dependence and with no
charm in the initial state. The four-flavor structure function
$F^{(4)}$ is instead computed by setting $m_c$ to zero and allowing
for charm in the initial state, and its associated PDF reabsorbs the
mass (collinear) divergences which are in turn resummed by means of
the DGLAP evolution. Finally, $F^{(3,0)}$ represents the massless
limit of $F^{(3)}$ where all the massive power corrections are set to
zero and only the logarithmically enhanced terms are retained. This
last term is meant to subtract the double counting terms resulting
from the sum of $F^{(3)}$ and $F^{(4)}$. In fact, the role of
$F^{(3,0)}$ is twofold: for $Q\gg m_c$, by definition $F^{(3)}$ and
$F^{(3,0)}$ tend to the same value so that the FONLL structure
function reduces to $F^{(4)}$. By contrast, in the region where
$Q \simeq m_c$ it can be shown that $F^{(d)}$ becomes subleading in
$\alpha_s$ reducing the FONLL structure function to $F^{(3)}$ up to
terms beyond the nominal perturbative accuracy.

It should be noticed that, even though $F^{(d)}$ in
eq.~(\ref{eq:FONLLdef}) becomes subleading in the low-energy region,
it might become numerically relevant and it is advisable to suppress
it. To this end, the term $F^{(d)}$ in eq.~(\ref{eq:FONLLdef}) is
usually replaced by:
\begin{equation}
F^{(d')}(x,Q,m_c) = D(Q,m_c)F^{(d)}(x,Q,m_c)\,,
\end{equation}
where the function $D(Q,m_c)$ is usually referred to as the
\textit{damping factor} and has the explicit form:
\begin{equation}\label{eq:dampingfactor}
D(Q,m_c)=\theta(Q^2-m_c^2)\left(1-\frac{m_c^2}{Q^2}\right)^2\,.
\end{equation}
The role of the damping factor is clearly that of setting $F^{(d')}$
to zero for $Q<m_c$, suppressing it for $Q\gtrsim m_c$, and reducing
it to $F^{(d)}$ for $Q\gg m_c$. It should be pointed out that the
particular functional form of the damping factor given in
eq.~(\ref{eq:dampingfactor}) is somewhat arbitrary. In fact, any
function $D$ such that $F^{(d')}$ and $F^{(d)}$ only differ by
power-suppressed terms, namely:
\begin{equation}
D(Q,m_c) = 1 + \mathcal{O}\left(\frac{m_c^2}{Q_2}\right)\,,
\end{equation}
is a formally suitable choice. In the results section we will also
consider the effect of varying the functional form of the damping
factor in order to estimate the associated theoretical uncertainty on
$m_c(m_c)$.

Given the possible different perturbative structure of the elements
that compose the FONLL structure function in eq.~(\ref{eq:FONLLdef}),
two possibilities for the definition of the perturbative ordering are
possible: the \textit{relative} and the \textit{absolute}
definitions. In the relative definition $F^{(4)}$ and $F^{(3)}$ are
combined using the same relative perturbative accuracy, that is LO
with LO, NLO with NLO, and so on. The absolute definition, instead, is
such that LO refers to $\mathcal{O}(\alpha_s^0)$ (parton model), NLO
to $\mathcal{O}(\alpha_s)$, and so forth. This issue is relevant in
the neutral-current case where the lowest non-vanishing order is
$\mathcal{O}(\alpha_s^0)$ for $F^{(4)}$ and $\mathcal{O}(\alpha_s)$
for $F^{(3)}$\footnote{This is strictly true only if the heavy-quark
  PDFs are dynamically generated via gluon splitting. In fact, the
  presence of an intrinsic heavy-quark component would introduce a
  $\mathcal{O}(\alpha_s^0)$ contribution also in $F^{(3)}$ leading to
  a ``realignment'' of the perturbative structure between $F^{(4)}$
  and $F^{(3)}$ (see Refs.~\cite{Ball:2015tna,Ball:2015dpa}).} such
that the relative and absolute orderings lead to different
prescriptions.

Beyond LO, there are currently three possible variants of the FONLL
scheme, all of them implemented in {\tt APFEL}:
\begin{itemize}
\item the FONLL-A variant adopts the absolute ordering at
  $\mathcal{O}(\alpha_s)$ and thus only terms up to this accuracy are
  included. This variant is formally NLO and thus also PDFs should be
  evolved using the same accuracy in the DGLAP evolution.
\item The FONLL-B variant is instead computed using the relative
  ordering at NLO. Therefore, $F^{(4)}$ is computed at
  $\mathcal{O}(\alpha_s)$ and combined with $F^{(3)}$ at
  $\mathcal{O}(\alpha_s^2)$. $F^{(3,0)}$ is instead computed dropping
  the non-logarithmic $\mathcal{O}(\alpha_s^2)$ term to match the
  accuracy of $F^{(4)}$ in the low-energy region. PDFs are again
  evolved at NLO.
\item Finally, the FONLL-C scheme adopts the absolute ordering at
  $\mathcal{O}(\alpha_s^2)$. This is formally a NNLO scheme thus PDFs
  should be evolved using the same accuracy.
\end{itemize}
Presently, no other variant beyond FONLL-C can be pursued because the
$\mathcal{O}(\alpha_s^3)$ massive coefficient functions are not known
yet. Approximate NNLO corrections valid near the partonic threshold,
in the high-energy (small-$x$) limit, and at high scales $Q^2 \gg m^2$
have been derived in Ref.~\cite{Kawamura:2012cr} and they are
currently employed by the ABM group to determine NNLO
PDFs~\cite{Alekhin:2013nda}.

As clear from the description above, the computations for the
three-flavor structure functions $F^{(3)}$ and $F^{(3,0)}$ depend
explicitly on the charm mass, while $F^{(4)}$ does not. In addition,
as already mentioned, the expressions needed to compute $F^{(3)}$ and
$F^{(3,0)}$ are usually given in terms of the pole mass. As a
consequence, one of the steps required to achieve a full formulation
of the FONLL structure functions in terms of $\overline{\rm MS}$
masses is the adaptation of the heavy-flavor contributions to the
structure functions. A thorough explanation of the procedure adopted
to perform such transformation can be found in
Ref.~\cite{Alekhin:2010sv} for both neutral- and charged-current
structure functions. In Sect.~\ref{sec:implementation} we re-derive
the main formulae and report the full expressions for the relevant
coefficient functions. It should be pointed out that the derivation
presented in Ref.~\cite{Alekhin:2010sv} is performed assuming
$\mu_R=m_c(m_c)$, $\mu_R$ being the renormalisation scale, and the
renormalisation scale dependence of $\alpha_s$ is restored only at the
end using the expansion of the solution of the relative RG
equation. Such a procedure implies that the heavy-quark mass is not
subject to the relative RG equation: in other words, the mass running
is not expressed explicitly.  The reason is that in the running of the
heavy-quark mass in $\overline{\rm MS}$ one can resum logarithms of
$\mu_R/m_c(m_c)$ and this is not required in a fixed-order
calculation. On the contrary, when dealing with a GM-VFN scheme like
FONLL, such a resummation is an important ingredient and thus should
be consistently incorporated into the derivation. For this reason, the
transition from pole to $\overline{\rm MS}$ masses of the massive
structure functions presented in Sect.~\ref{sec:implementation} is
done at the generic renormalisation scale $\mu_R$ and the connection
between $m_c(m_c)$ and $m_c(\mu_R)$ is established solving the
appropriate RG equation.

A further complication that arises in FONLL as a VFN scheme is the
fact that the involved running quantities, that is PDFs, $\alpha_s$
and the mass itself, have to be properly matched when crossing a
heavy-quark threshold in their evolution. The matching conditions for
PDFs and $\alpha_s$ are presently known up to
$\mathcal{O}(\alpha_s^2)$~\cite{Larin:1994va} and
$\mathcal{O}(\alpha_s^3)$~\cite{Chetyrkin:1997sg}, respectively, but
those for PDFs are given in terms of the pole mass. In the next
section we will show how to express them in terms of the
$\overline{\rm MS}$ mass up to the relevant accuracy. As far as the
matching of the mass is concerned, the expressions for the matching
conditions are given in Ref.~\cite{Chetyrkin:2000yt} up to
$\mathcal{O}(\alpha_s^3)$ also in terms of $\overline{\rm MS}$ mass.


\subsection{Implementation}\label{sec:implementation}

In this section we will describe in some detail the implementation of
the FONLL scheme in terms of the $\overline{\mbox{MS}}$ heavy-quark
masses in {\tt APFEL}.  Starting from the more usual definition of
structure functions in terms of pole masses, our goal is to
consistently replace them with the $\overline{\mbox{MS}}$ mass
definition.

\subsubsection{$\overline{\mbox{MS}}$ mass vs. pole mass}

The (scale independent) pole mass $M$ and the (scale dependent)
$\overline{\mbox{MS}}$ mass $m(\mu)$ arise from two different
renormalization procedures and, as already mentioned, in perturbation
theory they can be expressed one in terms of the other. The relation
connecting pole and $\overline{\mbox{MS}}$ mass definitions has been
computed in Ref.~\cite{Chetyrkin:1999qi} up to four loops. However, in
the following we will only need to go up to one loop and thus we
report here the corresponding relation:
\begin{equation}\label{eq:PoleToMSbar}
\frac{M}{m(\mu)} = 1 + h^{(1)}a_s+\mathcal{O}(a_s^2)\,,
\end{equation}
with:
\begin{equation}\label{eq:CoeffPoleToMSbar}
\begin{array}{l}
\displaystyle h^{(1)}(\mu,m(\mu)) = C_F\left(4 + 3L_{\mu m}\right)\,,
\end{array}
\end{equation}
where $C_F=4/3$ is one of the usual QCD color factors. Moreover, we
have defined:
\begin{equation}
a_s\equiv a_s(\mu) = \frac{\alpha_s(\mu)}{4\pi}\,,
\end{equation}
and:
\begin{equation}
L_{\mu m} = \ln\frac{\mu^2}{m^2(\mu)}\,.
\end{equation}
In the following we will use eq. (\ref{eq:PoleToMSbar}) to replace the
pole mass $M$ with the $\overline{\mbox{MS}}$ mass $m(\mu)$.

\subsubsection{Solution of the RGE for the running of the $\overline{\mbox{MS}}$ mass}

In order to evaluate the running of $m(\mu)$ with the renormalization
scale $\mu$ we have to solve the corresponding renormalization-group
equation (RGE):
\begin{equation}\label{eq:massRGE}
\mu^2\frac{dm}{d\mu^2} = m(\mu)\gamma_m(a_s) = -m(\mu)\sum_{n=0}^{\infty}\gamma_m^{(n)}a_s^{n+1}\,,
\end{equation}
whose first three coefficients can be taken from
Ref.~\cite{Chetyrkin:1999pq}\footnote{The following expressions have
  been adjusted taking into account our definition of $a_s$ which
  differs by a factor of 4 with respect to that of Ref.~\cite{Chetyrkin:1999pq}.}:
\begin{subequations} 
\begin{equation}
\gamma_m^{(0)} = 4 \,,
\end{equation}
\begin{equation}
\gamma_m^{(1)} = \frac{202}3 - \frac{20}{9}N_f\,,
\end{equation}
\begin{equation}
\gamma_m^{(2)} = 1249 - \left(\frac{2216}{27}+\frac{160}{3}\zeta_3\right)N_f-\frac{140}{81}N_f^2\,,
\end{equation}
\end{subequations} 
where $N_f$ is the number of active flavors. In addition, the RGE for
the running of $\alpha_s$ reads:
\begin{equation}\label{eq:alphasRGE}
\mu^2\frac{da_s}{d\mu^2} = \beta(a_s) = -\sum_{n=0}^{\infty}\beta_n a_s^{n+2}\,,
\end{equation}
with:
\begin{subequations} 
\begin{equation}
\beta_0 = 11-\frac23 N_f \,,
\end{equation}
\begin{equation}
\beta_1 = 102 - \frac{38}3 N_f\,.
\end{equation}
\begin{equation}
\beta_2 = \frac{2857}{2} - \frac{5033}{18}N_f + \frac{325}{54}N_f^2\,.
\end{equation}
\end{subequations} 
Combining eqs.~(\ref{eq:massRGE}) and~(\ref{eq:alphasRGE}) we obtain the
following differential equation:
\begin{equation}\label{runmass}
\frac{dm}{da_s} = \frac{\gamma_m(a_s)}{\beta(a_s)}m(a_s)\,,
\end{equation}
whose solution is:
\begin{equation}\label{eq:massRGEsolution}
m(\mu) = m(\mu_0)\exp\left[\int_{a_s(\mu_0)}^{a_s(\mu)}\frac{\gamma_m(a_s)}{\beta(a_s)}da_s\right]\,.
\end{equation}
In order to get an analytical expression out of
eq.~(\ref{eq:massRGEsolution}), one can expand the integrand in the
r.h.s. using the perturbative expansions of $\gamma_m(a_s)$ and
$\beta(a_s)$ given in eqs.~(\ref{eq:massRGE})
and~(\ref{eq:alphasRGE}). This allows us to solve the integral
analytically, obtaining:
\begin{equation}\label{eq:massRGEsolutionAnal}
\begin{array}{rcl}
m(\mu)&=&\displaystyle m(\mu_0)\left(\frac{a}{a_0}\right)^{c_0}\\
\\
&\times&\displaystyle
         \frac{1+(c_1-b_1c_0)a+\frac12[c_2-c_1b_1-b_2c_0+b_1^2c_0+(c_1-b_1c_0)^2]a^2}{1+(c_1-b_1c_0)a_0+\frac12[c_2-c_1b_1-b_2c_0+b_1^2c_0+(c_1-b_1c_0)^2]a_0^2}\,,
\end{array}
\end{equation}
where we have defined:
\begin{equation}\label{eq:SolCoefs}
b_i = \frac{\beta_i}{\beta_0}\quad\mbox{and}\quad c_i = \frac{\gamma_m^{(i)}}{\beta_0}\,,
\end{equation}
and $a\equiv a_s(\mu)$ and $a_0\equiv a_s(\mu_0)$.
Eq.~(\ref{eq:massRGEsolutionAnal}) represents the NNLO solution of the
RGE for the $\overline{\mbox{MS}}$ mass $m(\mu)$.

Of course, the NLO and the LO solutions can be easily extracted from
eq.~(\ref{eq:massRGEsolutionAnal}) just by disregarding the terms
proportional to $a^2$ and $a_0^2$ for the NLO solution and also the
terms proportional to $a$ and $a_0$ for the LO solution\footnote{In
  order to be consistent, the evaluation of $a$ and $a_0$
  eq.~(\ref{eq:massRGEsolutionAnal}) must be performed at the same
  perturbative order of $m(\mu)$. So, for instance, if one wants to
  evaluate the NNLO running of $m(\mu)$ also the value of $a$ and
  $a_0$ must be computed using the NNLO running.}.

\subsubsection{Matching conditions}\label{sec:MatchingConditions}

When working in the context of a VFN scheme, all running quantities
are often required to cross heavy-quark thresholds when evolving from
one scale to another. Such a transition in turn requires the matching
different factorization schemes whose content of active flavors
differs by one unit. In other words, if the perturbative evolution
leads from an energy region where (by definition) there are $N_f-1$
active flavors to another region where there are $N_f$ active flavors,
the two regions must be consistently connected and such a connection
can be evaluated perturbatively. This goes under the name of
\textit{matching conditions}.

In general, matching conditions give rise to discontinuities of the
running quantities at the matching scales and in the following we will
report the matching conditions up to NNLO in terms of the
$\overline{\mbox{MS}}$ heavy-quark thresholds for: $\alpha_s(\mu)$,
$m(\mu)$ and PDFs.

\subsubsection*{Matching of $\alpha_s(\mu)$}

The matching conditions for $\alpha_s$ were evaluated in
Ref.~\cite{Chetyrkin:1997sg} to three loops.  We report here the
relation up to two loops (again taking into account the factor 4
coming from the different definitions of $a$):
\begin{equation}\label{eq:alphaspole}
\frac{a^{(N_f-1)}(\mu)}{a^{(N_f)}(\mu)}=1-\frac23 L_{\mu M}a^{(N_f)}(\mu)+\left(\frac49L_{\mu M}^2-\frac{38}3L_{\mu M}-\frac{14}3\right)[a^{(N_f)}(\mu)]^2\,.
\end{equation}
$M$ being the pole mass of the $n$-th flavor. From
eq.~(\ref{eq:PoleToMSbar}) we can easily infer that:
\begin{equation}
\ln M^2 = \ln m^2(\mu) + 2\ln[1+h^{(1)}(\mu)a^{(N_f)}(\mu)]= \ln m^2(\mu) + 2h^{(1)}(\mu)a^{(N_f)}(\mu)+\mathcal{O}([a^{(N_f)}]^2)\,.
\end{equation}
Therefore, it is straightforward to see that:
\begin{equation}\label{eq:conversionLog}
L_{\mu M}  = L_{\mu m} - 2h^{(1)}a^{(N_f)}=L_{\mu m}-\left(\frac{32}3+8L_{\mu m}\right)a^{(N_f)}\,,
\end{equation}
so that:
\begin{equation}\label{eq:alphasmsbar}
\frac{a^{(N_f-1)}(\mu)}{a^{(N_f)}(\mu)}=1-\frac23 L_{\mu m}a^{(N_f)}(\mu)+\left(\frac49L_{\mu m}^2-\frac{22}3L_{\mu m}+\frac{22}9\right)[a^{(N_f)}(\mu)]^2\,,
\end{equation}
consistently with eq.~(20) of Ref.~\cite{Chetyrkin:2000yt}.

In order to simplify this expression, it is a common procedure to
perform the matching at the point where the logarithms vanish. In this
particular case, choosing $\mu=m(\mu)=m(m)$, we get:
\begin{equation}
a^{(N_f-1)}(m)=a^{(N_f)}(m)\left(1+\frac{22}9[a^{(N_f)}(m)]^2\right)\,,
\end{equation}
which can be easily inverted obtaining:
\begin{equation}\label{eq:alphasatthrs}
a^{(N_f)}(m)=a^{(N_f-1)}(m)\left(1-\frac{22}9[a^{(N_f-1)}(m)]^2\right)\,.
\end{equation}

It is interesting to observe that, in order to perform the matching as
described above, one just needs to know the value of $m(m)$. This is
the so-called RG-invariant $\overline{\mbox{MS}}$ mass.

\subsubsection*{Matching of $m(\mu)$}

The running of the $\overline{\mbox{MS}}$ masses also needs to be
matched at the heavy-quark thresholds. In particular, one needs to
match the $(N_f-1)$- with $(N_f)$-scheme for the mass $m_q(\mu)$, with
$q=c,b,t$, at the threshold $m_h(\mu)$, where $h=c,b,t$. From
Ref.~\cite{Chetyrkin:2000yt} we read:
\begin{equation}\label{eq:mqmc}
\frac{m_q^{(N_f-1)}(\mu)}{m_q^{(N_f)}(\mu)}=1+\left(\frac43L_{\mu m}^{(h)2}-\frac{20}9L_{\mu m}^{(h)}+\frac{89}{27}\right)[a^{(N_f)}(\mu)]^2\,,
\end{equation}
where:
\begin{equation}
L_{\mu m}^{(h)} =\ln\frac{\mu^2}{m_h^2(\mu)}\,.
\end{equation}
Exactly as before, if we choose to match the two schemes at the scale
$\mu=m_h(\mu)=m_h(m_h)$, the logarithmic terms vanish and we are left
with:
\begin{equation}\label{eq:MarchMhUp}
m_q^{(N_f-1)}(m_h)=\left(1+\frac{89}{27}[a^{(N_f)}(m_h)]^2\right)m_q^{(N_f)}(m_h)\,,
\end{equation}
whose inverse is:
\begin{equation}\label{eq:MarchMhDown}
m_q^{(N_f)}(m_h)=\left(1-\frac{89}{27}[a^{(N_f-1)}(m_h)]^2\right)m_q^{(N_f-1)}(m_h)\,.
\end{equation}

\subsubsection*{Matching of PDFs}

To conclude the section on the matching conditions, we finally
consider PDFs. One can write the singlet and the gluon in the
$(N_f)$-scheme in terms of singlet and gluon in the $(N_f-1)$-scheme
at any scale $\mu$ as follows:
\begin{equation}\label{eq:MatchPDFsPole}
\begin{array}{c}
\displaystyle {\Sigma^{(N_f)} \choose g^{(N_f)}}=\begin{pmatrix}1+a_s^2[A_{qq,h}^{N\!S,(2)}+\tilde{A}^{S,(2)}_{hq}] & a_s\tilde{A}^{S,(1)}_{hg}+a_s^2\tilde{A}^{S,(2)}_{hg}\\
a_s^2A^{S,(2)}_{gq,h} & 1+a_sA_{gg,h}^{S,(1)}+a_s^2A_{gg,h}^{S,(2)}\end{pmatrix}{\Sigma^{(N_f-1)} \choose g^{(N_f-1)}}\,,
\end{array}
\end{equation}
where the form of the functions entering the transformation matrix
above are given in Appendix B of Ref.~\cite{Buza:1996wv} in terms of
the pole mass. We omit the matching conditions for the non-singlet PDF
combinations because they have no $\mathcal{O}(a_s)$ correction and
the first correction appears at $\mathcal{O}(a_s^2)$. This leaves the
conversion from the pole to the $\overline{\mbox{MS}}$ mass scheme
unaffected up to NNLO.

In order to replace the pole mass $M$ with the $\overline{\mbox{MS}}$
mass $m(\mu)$, we just have to plug eq.~(\ref{eq:conversionLog}) into
eq.~(\ref{eq:MatchPDFsPole}). In doing so, only the $\mathcal{O}(a_s)$
terms proportional to $\ln(\mu^2/M^2)$ play a role in the conversion
up to NNLO. Since the functions $\tilde{A}^{S,(1)}_{hg}$ and
$A_{gg,h}^{S,(1)}$ can be written as:
\begin{equation}\label{eq:OasMatchPDFs}
\begin{array}{l}
\displaystyle \tilde{A}^{S,(1)}_{hg}\left(x,\frac{\mu^2}{M^2}\right) = f_1(x)\ln\frac{\mu^2}{M^2}\,,\\
\\
\displaystyle A^{S,(1)}_{gg,h}\left(x,\frac{\mu^2}{M^2}\right) = f_2(x)\ln\frac{\mu^2}{M^2}
\end{array}\,,
\end{equation}
where:
\begin{equation}\label{eq:OasPDFsMatchCoeff}
\begin{array}{l}
\displaystyle f_1(x)= 4 T_R[x^2+(1-x)^2]\,,\\
\\
\displaystyle f_2(x)= -\frac43 T_R \delta(1-x)\,,
\end{array}
\end{equation}
replacing $M$ with $m$ in eq.~(\ref{eq:OasMatchPDFs}) using
eq. (\ref{eq:conversionLog}) leads to:
\begin{equation}
\begin{array}{l}
\displaystyle \tilde{A}^{S,(1)}_{hg}\left(x,\frac{\mu^2}{m^2}\right) = f_1(x)\ln\frac{\mu^2}{m^2}-2h^{(1)}(\mu)f_1(x)a_s(\mu)\,,\\
\\
\displaystyle A^{S,(1)}_{gg,h}\left(x,\frac{\mu^2}{m^2}\right) = f_2(x)\ln\frac{\mu^2}{m^2}-2h^{(1)}(\mu)f_2(x)a_s(\mu)\,.
\end{array}
\end{equation}
Therefore eq. (\ref{eq:MatchPDFsPole}) in terms of $m$ becomes:
\begin{equation}
\begin{array}{c}
\displaystyle {\Sigma^{(N_f)}\choose g^{(N_f)}}=\begin{pmatrix}1+a_s^2[A_{qq,h}^{N\!S,(2)}+\tilde{A}^{S,(2)}_{hq}] & a_s\tilde{A}^{S,(1)}_{hg}+a_s^2[\tilde{A}^{S,(2)}_{hg}-2h^{(1)}f_1]\,,\\
a_s^2A^{S,(2)}_{gq,h} & 1+a_sA_{gg,h}^{S,(1)}+a_s^2[A_{gg,h}^{S,(2)}-2h^{(1)}f_2]\end{pmatrix}{\Sigma^{(N_f-1)} \choose g^{(N_f-1)}}\,.
\end{array}
\end{equation}

As usual, we choose to match the $(N_f)$-scheme to the
$(N_f-1)$-scheme at $\mu = m(\mu) = m(m)$ so that all the logarithmic
terms vanish, obtaining:
\begin{equation}
\begin{array}{c}
\displaystyle {\Sigma^{(N_f)}\choose g^{(N_f)}}=\begin{pmatrix}1+a_s^2[A_{qq,h}^{N\!S,(2)}+\tilde{A}^{S,(2)}_{hq}] & a_s^2[\tilde{A}^{S,(2)}_{hg}-2h^{(1)}f_1]\\
a_s^2A^{S,(2)}_{gq,h} & 1+a_s^2[A_{gg,h}^{S,(2)}-2h^{(1)}f_2]\end{pmatrix}{\Sigma^{(N_f-1)} \choose g^{(N_f-1)}}\,.
\end{array}
\end{equation}

\subsubsection*{Renormalization scale variation}

The scale $\mu$ that appears in $a_s$ and $m_q$ is the
\textit{renormalization} scale, which we will now denote as $\mu_R$.
The scale that explicitly appears in the PDFs is instead the
\textit{factorization} scale, which we will now denote with
$\mu_F$. In principle, renormalization and factorization scales are
different but one usually takes them to be proportional to each other,
as $\mu_R = \kappa \mu_F$, where $\kappa$ can be any real
number\footnote{It should be noticed that in the case $\kappa\neq1$
  PDFs acquire an implicit dependence on $\mu_R$ that comes from a
  redefinition of the splitting functions that in turn derives from
  the expansion of $\alpha_s(\mu_R)$ around $\mu_R=\mu_F$.}.

The most common choice when matching the $(N_f-1)$-scheme to the
$(N_f)$-scheme is to set $\mu_F$ equal to heavy-quark thresholds
($M_c$, $M_b$ and $M_t$ in the pole-mass scheme and $m_c(m_c)$,
$m_b(m_b)$ and $m_t(m_t)$ in the $\overline{\mbox{MS}}$ scheme). In
doing so, the logarithmic terms in the PDF matching conditions are
assured to vanish.  However, if $\kappa$ is different from one, the
logarithmic terms in the matching conditions for $a_s(\mu_R)$ and
$m_q(\mu_R)$ do not vanish anymore. In the following we will show how
the matching conditions for $a_s$ and $m_q$ change for $\kappa\neq1$.

Let us start with $\alpha_s$. Inverting eq.~(\ref{eq:alphasmsbar}) we
obtain:
\begin{equation}
\frac{a^{(N_f)}(\mu_R)}{a^{(N_f-1)}(\mu_R)} = 1 + c_1a^{(N_f-1)}(\mu_R) + c_2 [a^{(N_f-1)}(\mu_R)]^2\,,
\end{equation}
where:
\begin{equation}
c_1 = \frac23 L_{\mu m}
\quad\mbox{and}\quad
c_2 = \frac49L_{\mu m}^2+\frac{22}3L_{\mu m}-\frac{22}9\,.
\end{equation}
Setting $\mu_F=\kappa \mu_F$, we have that:
\begin{equation}
L_{\mu m} = \ln\frac{\mu_R}{m(\mu_R)}=\ln\frac{\kappa\mu_F}{m(\kappa \mu_F)}\,.
\end{equation}
As usual, the matching scale is chosen to be $\mu_F = m(m)$, so that:
\begin{equation}
L_{\mu m} \rightarrow \ln\kappa + \ln\frac{m(m)}{m(\kappa m)}\,.
\end{equation}
But using eq.~(\ref{eq:massRGEsolution}), it is easy to see that:
\begin{equation}
 \ln\frac{m(m)}{m(\kappa m)}=a_s(\kappa m)\gamma_m^{(0)}\ln\kappa+\mathcal{O}[a_s^2(\kappa m)]\,,
\end{equation}
so that:
\begin{equation}\label{eq:LmumExp}
L_{\mu m} \rightarrow [1+\gamma_m^{(0)}a_s(\kappa m)]\ln\kappa\,.
\end{equation}
It should be noticed that in the eq.~(\ref{eq:LmumExp}), since
$a_s^{(N_f-1)}=a_s^{(N_f)}+\mathcal{O}([a_s^{(N_f)}]^2)$, it does not
matter whether one uses $a_s^{(N_f)}(\kappa m)$ or
$a_s^{(N_f-1)}(\kappa m)$ because the difference would be subleading
up to NNLO.

Therefore, setting $\mu=\mu_R=\kappa m(m) = \kappa m$ in
eq.~(\ref{eq:alphasmsbar}) and using eq. (\ref{eq:LmumExp}), one gets:
\begin{equation}
\begin{array}{rcl}
a^{(N_f-1)}(\kappa m)&=&\displaystyle a^{(N_f)}(\kappa m)\bigg\{1-\frac23
  \ln\kappa\,a^{(N_f)}(\kappa
  m)\\
\\
&+&\displaystyle \left[\frac49\ln^2\kappa-\frac{2}3\left(\gamma_m^{(0)}+11\right)\ln\kappa+\frac{22}9\right][a^{(N_f)}(\kappa
  m)]^2\bigg\}\,,
\end{array}
\end{equation}
whose inverse is:
\begin{equation}
\begin{array}{rcl}
a^{(N_f)}(\kappa m)&=&\displaystyle a^{(N_f-1)}(\kappa m)\bigg\{1+\frac23
  \ln\kappa\,a^{(N_f-1)}(\kappa
  m)\\
\\
&+&\displaystyle \left[\frac49\ln^2\kappa+\frac{2}3\left(\gamma_m^{(0)}+11\right)\ln\kappa-\frac{22}9\right][a^{(N_f-1)}(\kappa
  m)]^2\bigg\}\,.
\end{array}
\end{equation}

Now let us turn to $m_q$. In this case there is not much to do. In
fact, for an arbitrary matching point the matching condition of the
$\overline{\mbox{MS}}$ mass starts at $\mathcal{O}(\alpha_s^2)$
(cfr. eq. (\ref{eq:mqmc})), therefore writing $L_{\mu m}$ in terms of
$\ln\kappa$ would give rise to subleading terms up to NNLO (see
eq. (\ref{eq:LmumExp})). As a consequence, we have that:
\begin{equation}
m_q^{(N_f-1)}(\kappa m_h)=\left[1+\left(\frac43\ln^2\kappa-\frac{20}9\ln\kappa+\frac{89}{27}\right)[a^{(N_f)}(\kappa m_h)]^2\right]m_q^{(N_f)}(\kappa m_h)\,,
\end{equation}
whose inverse is:
\begin{equation}
m_q^{(N_f)}(\kappa
m_h)=\left[1-\left(\frac43\ln^2\kappa-\frac{20}9\ln\kappa+\frac{89}{27}\right)[a^{(N_f-1)}(\kappa
  m_h)]^2\right]m_q^{(N_f-1)}(\kappa m_h)\,.
\end{equation}

\subsubsection{Structure functions}

We finally turn to discuss how the DIS massive structure functions
change when expressing them in terms of the $\overline{\mbox{MS}}$
masses. We will first consider the neutral-current (NC) massive
structure functions up to $\mathcal{O}(\alpha_s^2)$, which is the
highest perturbative order at which corrections are known exactly, and
then we will consider the charged-current (CC) massive structure
functions again up to the highest perturbative order exactly
known\footnote{In a recent publication~\cite{Berger:2016inr} the
  $\mathcal{O}(\alpha_s^2)$ corrections (NNLO) to charm production in
  CC DIS were presented. However, no analytical expression was
  provided.}, that is $\mathcal{O}(\alpha_s)$. In order to shorten the
notation, we will adopt the following definitions:
$$
M =\;\mbox{pole mass},\quad m\equiv m(\mu) =\;\overline{\mbox{MS}}\mbox{ mass},\quad a_s\equiv a_s(\mu),\quad h^{(l)}\equiv h^{(l)}(\mu,m(\mu))\,.
$$

\subsubsection*{Neutral current}

Dropping all the unnecessary dependences, the NC massive structure
functions up to $\mathcal{O}(a_s^2)$ have the form:
\begin{equation}
F = a_sF^{(0)}(M) + a_s^2F^{(1)}(M) + \mathcal{O}(a_s^3)\,.
\end{equation}
The goal is to replace explicitly the pole mass $M$ with the
$\overline{\mbox{MS}}$ mass $m$ using eq.~(\ref{eq:PoleToMSbar}). To
this end, following the procedure adopted in
Ref.~\cite{Alekhin:2010sv}, we expand $F^{(0)}(M)$ and $F^{(1)}(M)$
around $M=m$:
\begin{equation}
F^{(l)}(M) = \sum_{n=0}^{\infty}\frac1{n!}\frac{d^n F^{(l)}}{dM^n}\bigg|_{M=m}(M-m)^n\,,
\end{equation}
so that, up to $\mathcal{O}(a_s^2)$, what we need is:
\begin{equation}
\begin{array}{l}
\displaystyle F^{(0)}(M) = F^{(0)}(m) + a_smh^{(1)}\frac{dF^{(0)}}{dM}\bigg|_{M=m}\,,\\
\\
\displaystyle F^{(1)}(M) = F^{(1)}(m)\,.
\end{array}
\end{equation}
Finally, we have that:
\begin{equation}\label{eq:changescheme}
F = a_sF^{(0)}(m) + a_s^2\left[F^{(1)}(m)+mh^{(1)}\frac{dF^{(0)}}{dM}\bigg|_{M=m}\right]\,.
\end{equation}

We now need to evaluate explicitly the derivative in
eq. (\ref{eq:changescheme}). First of all we observe that:
\begin{equation}\label{eq:convolution0}
F^{(0)}(M) = x\int_x^{x_{\mbox{\tiny max}}(M)} \frac{dz}{z}g\left(\frac{x}{z}\right)C_g^{(0)}(\eta(z,M),\xi(M),\chi(M))\,,
\end{equation}
where $g$ is the gluon distribution and we have used the following
definitions:
\begin{equation}
x_{\mbox{\tiny max}}(M)=\frac1{1+\frac{4M^2}{Q^2}},\quad\eta(z,M) = \frac{Q^2}{4M^2}\left(\frac1z-1\right)-1,\quad \xi(M) =\frac{Q^2}{M^2},\quad \chi(M) =\frac{\mu^2}{M^2}\,.
\end{equation}
Defining:
\begin{equation}
G(z,M)=\frac{x}{z}g\left(\frac{x}{z}\right)C_g^{(0)}(\eta(z,M),\xi(M),\chi(M))\,,
\end{equation}
the derivative of eq.~(\ref{eq:convolution0}) can be written as:
\begin{equation}
\frac{dF^{(0)}}{dM} = \frac{d}{dM}\int_x^{x_{\mbox{\tiny max}}(M)} dzG(z,M) = \frac{d\widetilde{G}(x_{\mbox{\tiny max}}(M),M)}{dM}-\frac{d\widetilde{G}(x,M)}{dM}\,,
\end{equation}
where $\widetilde{G}(z,M)$ is the primitive of $G(z,M)$ with respect to $z$ (i.e. $\partial\widetilde{G}/\partial z = G$). But:
\begin{equation}
\frac{d\widetilde{G}(x_{\mbox{\tiny max}}(M),M)}{dM} = \frac{d \widetilde{G}(x_{\mbox{\tiny max}},M)}{d M}+\frac{dx_{\mbox{\tiny max}}}{dM}G(x_{\mbox{\tiny max}},M)\,,
\end{equation}
thus:
\begin{equation}\label{eq:withbound}
\begin{array}{c}
\displaystyle \frac{dF^{(0)}}{dM} = \frac{\partial \widetilde{G}(x_{\mbox{\tiny max}},M)}{\partial M}-\frac{\partial\widetilde{G}(x,M)}{\partial M}+\frac{dx_{\mbox{\tiny max}}}{dM}G(x_{\mbox{\tiny max}},M) =\\
\\
\displaystyle \int_x^{x_{\mbox{\tiny max}}(M)} dz\frac{\partial G(z,M)}{\partial M}+\frac{dx_{\mbox{\tiny max}}}{dM}G(x_{\mbox{\tiny max}},M)\,.
\end{array}
\end{equation}
It can be shown that the boundary term in eq.~(\ref{eq:withbound})
vanishes (see Ref.~\cite{Alekhin:2010sv}), thus it can be omitted.

Gathering all pieces and taking into account that:
\begin{equation}
\frac{\partial G(z,M)}{\partial M} = \frac{x}{z}g\left(\frac{x}{z}\right)\frac{\partial C_g^{(0)}}{\partial M}\,,
\end{equation}
we have that:
\begin{equation}\label{eq:withoutbound}
\begin{array}{rcl}
\displaystyle \frac{dF^{(0)}}{dM}\bigg|_{M=m}&=&\displaystyle
                                                 \left[x\int_x^{x_{\mbox{\tiny
                                                 max}}(M)}\frac{dz}{z}g\left(\frac{x}{z}\right)\frac{\partial
                                                 C_g^{(0)}}{\partial
                                                 M}\right]\Bigg|_{M=m}\\
\\
&=&\displaystyle x\int_x^{x_{\mbox{\tiny
    max}}(m)}\frac{dz}{z}g\left(\frac{x}{z}\right)\left[\frac{\partial
    C_g^{(0)}}{\partial M}\right]\Bigg|_{M=m}\,.
\end{array}
\end{equation}

Finally, considering that:
\begin{equation}
F^{(1)}(M) = \sum_{i=q,\overline{q},g}x\int_x^{x_{\mbox{\tiny max}}(M)} \frac{dz}{z}q_i\left(\frac{x}{z}\right)C_i^{(1)}(z,M)
\end{equation}
and using eqs.~(\ref{eq:changescheme}) and~(\ref{eq:withoutbound}),
one can explicitly write down the full structure of the massive
structure functions ($F_2$ and $F_L$) in terms of
$\overline{\mbox{MS}}$ masses up to $\mathcal{O}(\alpha_s^2)$ as
follows:
\begin{equation}\label{eq:masterNC}
\begin{array}{c}
\displaystyle F = x\int_x^{x_{\mbox{\tiny max}}(m)} \frac{dz}{z}g\left(\frac{x}{z}\right)\left[a_sC_g^{(0)}(z,m)+a_s^2\left(C_g^{(1)}(z,m)+mh^{(1)}\left[\frac{\partial C_g^{(0)}}{\partial M}\right]\Bigg|_{M=m}\right)\right]+\\
\\
\displaystyle \sum_{i=q,\overline{q}}x\int_x^{x_{\mbox{\tiny max}}(M)} \frac{dz}{z}q_i\left(\frac{x}{z}\right)a_s^2C_i^{(1)}(z,M)\,.
\end{array}
\end{equation}

In order to carry out the implementation, we need to evaluate
explicitly the derivative of $C_g^{(0)}$ in eq. (\ref{eq:masterNC})
and this must be done separately for $F_2$ and $F_L$.

We consider $F_2$ first. The explicit expression of $C_{2,g}^{(0)}$ is
the following:
\begin{equation}
\begin{array}{rl}
\displaystyle C_{2,g}^{(0)}(z,Q^2,M^2)=&\displaystyle T_R\Big\{2(1-6\epsilon-4\epsilon^2)I_2(\epsilon,z)-2(1-2\epsilon)I_1(\epsilon,z)+I_0(\epsilon,z)+\\
\\
& \displaystyle -4(2-\epsilon)J_2(\epsilon,z)+4(2-\epsilon)J_1(\epsilon,z)-J_0(\epsilon,z)\Big\}\,,
\end{array}
\end{equation}
where:
\begin{equation}\label{eq:iq}
I_q(\epsilon,z) = z^q\ln\left(\frac{1+v}{1-v}\right)\,.
\end{equation}
\begin{equation}\label{eq:jq}
J_q(\epsilon,x) = z^q v\,,
\end{equation}
with:
\begin{equation}\label{eq:definitions1}
\epsilon = \frac{M^2}{Q^2}\,,\quad
a=\frac1{1+4\epsilon}\quad\mbox{and}\quad v=\sqrt{1-4\epsilon\frac{z}{1-z}}\,.
\end{equation}
From the definitions in eq.~(\ref{eq:definitions1}), we obtain:
\begin{equation}\label{eq:derivatives}
\begin{array}{rl}
\displaystyle \frac{\partial}{\partial M} &\displaystyle = \frac{\partial \epsilon}{\partial M} \frac{\partial}{\partial \epsilon} = \frac{2\epsilon}{M}\frac{\partial}{\partial \epsilon}\,,\\
\\
\displaystyle \frac{\partial}{\partial M} &\displaystyle = \frac{\partial \epsilon}{\partial M} \frac{\partial v}{\partial \epsilon} \frac{\partial}{\partial v} = -\frac{1-v^2}{Mv}\frac{\partial}{\partial v}\,.
\end{array}
\end{equation}
Therefore:
\begin{equation}
\begin{array}{rl}
\displaystyle \frac{\partial C_{2,g}^{(0)}}{\partial M}=&\displaystyle \frac{1}{M}T_R\Bigg\{2\epsilon\Big[2(-6-8\epsilon)I_2+4I_1+4J_2-4J_1\Big]\\
\\
&\displaystyle -\frac{1-v^2}{v}\Bigg[2(1-6\epsilon-4\epsilon^2)\frac{\partial I_2}{\partial v}-2(1-2\epsilon)\frac{\partial I_1}{\partial v}+\frac{\partial I_0}{\partial v}\\
\\
& \displaystyle -4(2-\epsilon)\frac{\partial J_2}{\partial v}+4(2-\epsilon)\frac{\partial J_1}{\partial v}-\frac{\partial J_0}{\partial v}\Bigg]\Bigg\}\,.
\end{array}
\end{equation}
To find the explicit expression, we just need to evaluate the
derivative of $I_q$ and $J_q$ starting from eqs. (\ref{eq:iq}) and
(\ref{eq:jq}) which is easily done:
\begin{equation}
\begin{array}{rcl}
\displaystyle \frac{\partial I_q}{\partial v} &=& \displaystyle\frac{2 z^q}{1-v^2}\,,\\
\\
\displaystyle \frac{\partial J_q}{\partial v} &=& \displaystyle z^q\,.
\end{array}
\end{equation}
In the end we get:
\begin{equation}\label{eq:dC2g0}
\begin{array}{rl}
\displaystyle \frac{\partial C_{2,g}^{(0)}}{\partial M}=&\displaystyle
                                                          \frac{1}{M}T_R\Bigg\{4\epsilon\left[(-6-8\epsilon)z^2+2z\right]\ln\left(\frac{1+v}{1-v}\right)+8\epsilon
                                                          z(z-1)v\\
\\
&\displaystyle -\frac{2}{v}\left[2(1-6\epsilon-4\epsilon^2)z^2-2(1-2\epsilon)z+1\right]\\
\\
& \displaystyle -\frac{1-v^2}{v}\left[-4(2-\epsilon)z^2+4(2-\epsilon)z-1\right]\Bigg\}\,.
\end{array}
\end{equation}

The implementation of the FONLL scheme given in
eq.~(\ref{eq:FONLLdef}) requires the massless limit of the massive
structure functions. In practice this means that one needs to compute
the limit $M\rightarrow 0$ of the massive coefficient functions
retaining the logarithmic enhanced terms. In order to apply this
recipe to eq.~(\ref{eq:dC2g0}), we observe that:
\begin{equation}
\epsilon \mathop{\longrightarrow}_{M\rightarrow 0} 0 \,,\quad v \mathop{\longrightarrow}_{M\rightarrow 0} 1\,,
\end{equation}
and that:
\begin{equation}
\ln\left(\frac{1+v}{1-v}\right) \mathop{\longrightarrow}_{M\rightarrow
  0} \ln\frac{Q^2(1-z)}{M^2z}\,,
\end{equation}
so that:
\begin{equation}\label{eq:dC2g00}
\displaystyle \frac{\partial C_{2,g}^{(0)}}{\partial M}\mathop{\longrightarrow}_{M\rightarrow 0}\displaystyle \frac{\partial C_{2,g}^{0,(0)}}{\partial M}=
                                                          -\frac{2}{M}T_R\left(2z^2-2z+1\right)\,.
\end{equation}

We now turn to consider $F_L$. In this case the the gluon coefficient
function takes the simpler form:
\begin{equation}
C_{L,g}^{(0)}\left(z,Q^2,M^2\right)= T_R\left[-8\epsilon I_2(\epsilon,z)-4J_2(\epsilon,z)+4J_1(\epsilon,z)\right]\,.
\end{equation}
Therefore, using eq. (\ref{eq:derivatives}), we immediately get:
\begin{equation}
\frac{\partial C_{L,g}^{(0)}}{\partial M} = \frac1{M}T_R\left[-16\epsilon
    z^2\ln\left(\frac{1+v}{1-v}\right) +\frac{8\epsilon z^2}{v}-\frac{1-v^2}{v}\left(-4z^2+4z\right)\right]\,.
\end{equation}
It is finally easy to realize that:
\begin{equation}
\frac{\partial C_{L,g}^{(0)}}{\partial
  M}\mathop{\longrightarrow}_{M\rightarrow 0} \frac{\partial C_{L,g}^{0,(0)}}{\partial M}=0\,.
\end{equation}

\subsubsection*{Charged current}

In this section we consider the CC massive structure functions. The
treatment follows the exact same steps as the NC structure functions,
with the only difference being that in the CC case the first
non-vanishing term is $\mathcal{O}(a_s^0)$. This means that,
truncating the perturbative expansion at $\mathcal{O}(a_s)$, we have:
\begin{equation}
F_k = F_k^{(0)}(M) + a_sF_k^{(1)}(M) + \mathcal{O}(a_s^2)\,,
\end{equation}
with $k=2,3,L$. Therefore, expanding $F^{(0)}$ and $F^{(1)}$ around $M=m$ and keeping only the terms up to $\mathcal{O}(a_s)$, one obtains:
\begin{equation}\label{eq:expCC}
F_k = F_k^{(0)}(m) + a_s\left[F_k^{(1)}(m)+mh^{(1)}\frac{dF_k^{(0)}}{dM}\bigg|_{M=m}\right]\,.
\end{equation}

The leading-order contribution can be written as follows:
\begin{equation}
F^{(0)}_k(M) = b_k(M)s'(\xi(M))\,,
\end{equation}
where:
\begin{equation}\label{eq:definitions}
\xi = x\underbrace{\left(1+\frac{M^2}{Q^2}\right)}_{\frac1\lambda}=\frac{x}\lambda\quad\mbox{and}\quad
\left\{\begin{array}{l}
b_2 = \xi\\
b_3 = 1\\
b_L = (1-\lambda)\xi
\end{array}
\right.\,,
\end{equation}
where we have also defined:
\begin{equation}
s' = 2|V_{cs}|^2s+2|V_{cd}|^2d\,.
\end{equation}
Therefore:
\begin{equation}\label{eq:derivCC}
\begin{array}{rcl}
\displaystyle mh^{(1)}\frac{dF^{(0)}_k}{dM}\bigg|_{M=m} &=&\displaystyle
  mh^{(1)}\frac{d\xi}{dM}\frac{dF^{(0)}_k}{d\xi}\bigg|_{M=m} \\
\\
&=&\displaystyle 
  2h^{(1)}(1-\lambda)\xi\left[\frac{db_k}{d\xi}s'(\xi)+b_k(\xi)\frac{ds'}{d\xi}\right]\bigg|_{M=m}\,,
\end{array}
\end{equation}
that can be conveniently rewritten as:
\begin{equation}
mh^{(1)}\frac{dF^{(0)}_k}{dM}\bigg|_{M=m} = 2h^{(1)}(1-\lambda)\left[\left(\frac{db_k}{d\xi}-\frac{b_k}{\xi}\right)+b_k(\xi)\frac{d}{d\xi}\right]\xi s'(\xi)\bigg|_{M=m}\,,
\end{equation}
so that, using eq.~(\ref{eq:definitions}), we have that:
\begin{equation}\label{eq:CCcorrections}
\begin{array}{rcl}
\displaystyle mh^{(1)}\frac{dF^{(0)}_2}{dM}\bigg|_{M=m} &=& \displaystyle
2h^{(1)}(1-\lambda)\xi\frac{d}{d\xi} \xi s'(\xi)\bigg|_{M=m}\,,\\
\\
\displaystyle mh^{(1)}\frac{dF^{(0)}_3}{dM}\bigg|_{M=m} &=& \displaystyle
2h^{(1)}(1-\lambda)\frac{1}{\xi}\left[ \xi\frac{d}{d\xi}-1\right]\xi
s'(\xi)\bigg|_{M=m}\,,\\
\\
\displaystyle mh^{(1)}\frac{dF^{(0)}_L}{dM}\bigg|_{M=m} &=& \displaystyle
2h^{(1)}(1-\lambda)^2\xi \frac{d}{d\xi}\xi s'(\xi)\bigg|_{M=m}\,.
\end{array}
\end{equation}

Finally, we notice that in the massless limit, where
$\lambda\rightarrow 1$, all expressions in
eq.~(\ref{eq:CCcorrections}) vanish, with the consequence that the CC
massive structure functions up to $\mathcal{O}(a_s)$ in terms of the
pole mass $M$ or the $\overline{\mbox{MS}}$ mass $m$ are exactly the
same.

\subsection{Benchmark}

In order to validate the implementation in {\tt APFEL}, we have
benchmarked it against public codes. To the best of our knowledge,
there exist no public codes able to compute structure functions in the
FONLL scheme with $\overline{\rm MS}$ masses. For this reason the best
we could do is to benchmark the various ingredients separately.

As a first step, we present the benchmark of the running of PDFs,
$\alpha_s$ and $m_c$\footnote{The running of $m_b$ and $m_t$ has also
  been checked finding the same lavel of accuracy found for $m_c$.} in
the VFN scheme with $\overline{\rm MS}$ heavy-quark thresholds. The
difference with respect to the more common pole-mass formulation
arises from the fact that the matching of the evolutions at the
heavy-quark thresholds needs to be adapted to take into account the
different scheme used to renormalize the masses. The full set of such
matching conditions for PDFs, $\alpha_s$ and $m_c$ has been collected
in Sect.~\ref{sec:implementation}.

\begin{figure}
  \begin{center}
    \includegraphics[width=0.65\textwidth]{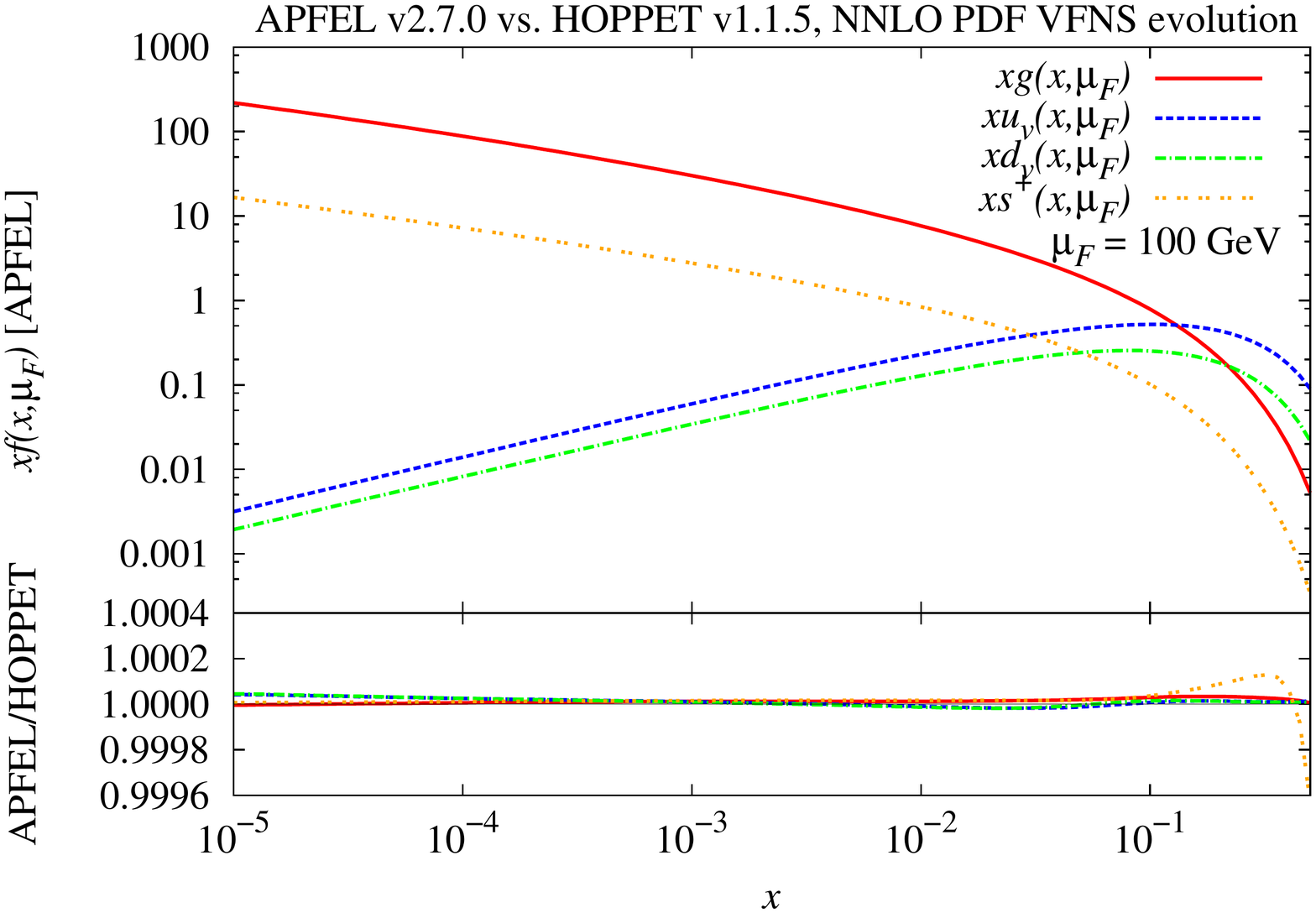}
  \end{center}
  \caption{\label{fig:APFELvsHOPPET} Comparison between {\tt APFEL}
    v2.7.0 and {\tt HOPPET} v1.1.5 for the VFNS DGLAP evolution at
    NNLO with $\overline{\rm MS}$ heavy-quark thresholds. The
    evolution settings, $i.e.$ initial scale PDFs, reference value of
    $\alpha_s$, and heavy-quark thresholds, are the same as used in
    the Les Houches PDF evolution benchmark~\cite{Dittmar:2005ed}. The
    upper inset shows the gluon PDF $xg$, the valence up and down PDFs
    $xu_v \equiv xu - x\overline{u}$ and
    $xd_v \equiv xd - x\overline{d}$, respectively, and the total
    strangeness $xs^+ \equiv xs + x\overline{s}$ at $\mu_F = 100$ GeV
    as functions of the Bjorken variable $x$ as returned by {\tt
      APFEL}. In the lower inset the ratio to {\tt HOPPET} is
    displayed showing a relative difference of $10^{-4}$ or better all
    over the considered range.}
\end{figure}

We start with the DGLAP PDF evolution in the VFN scheme with
$\overline{\rm MS}$ heavy-quark thresholds. A careful benchmark was
already presented in the original {\tt APFEL} publication. In
particular, the {\tt APFEL} evolution has been checked against the
{\tt HOPPET} code~\cite{Salam:2008qg} v1.1.5, finding a very good
agreement at the $\mathcal{O}\left(10^{-4}\right)$ level or
better. Since then, {\tt APFEL} has undergone several changes and
improvements and thus we repeated the benchmark using the same
settings and finding the same level of agreement with {\tt HOPPET}, as
shown in Fig.~\ref{fig:APFELvsHOPPET} for a representative set of
combinations of PDFs\footnote{We observe that, thanks to a better
  interpolation strategy, the predictions at the transition regions
  between internal $x$-space subgrids is now smoother.}.

\begin{figure}
  \begin{center}
    \includegraphics[width=0.49\textwidth]{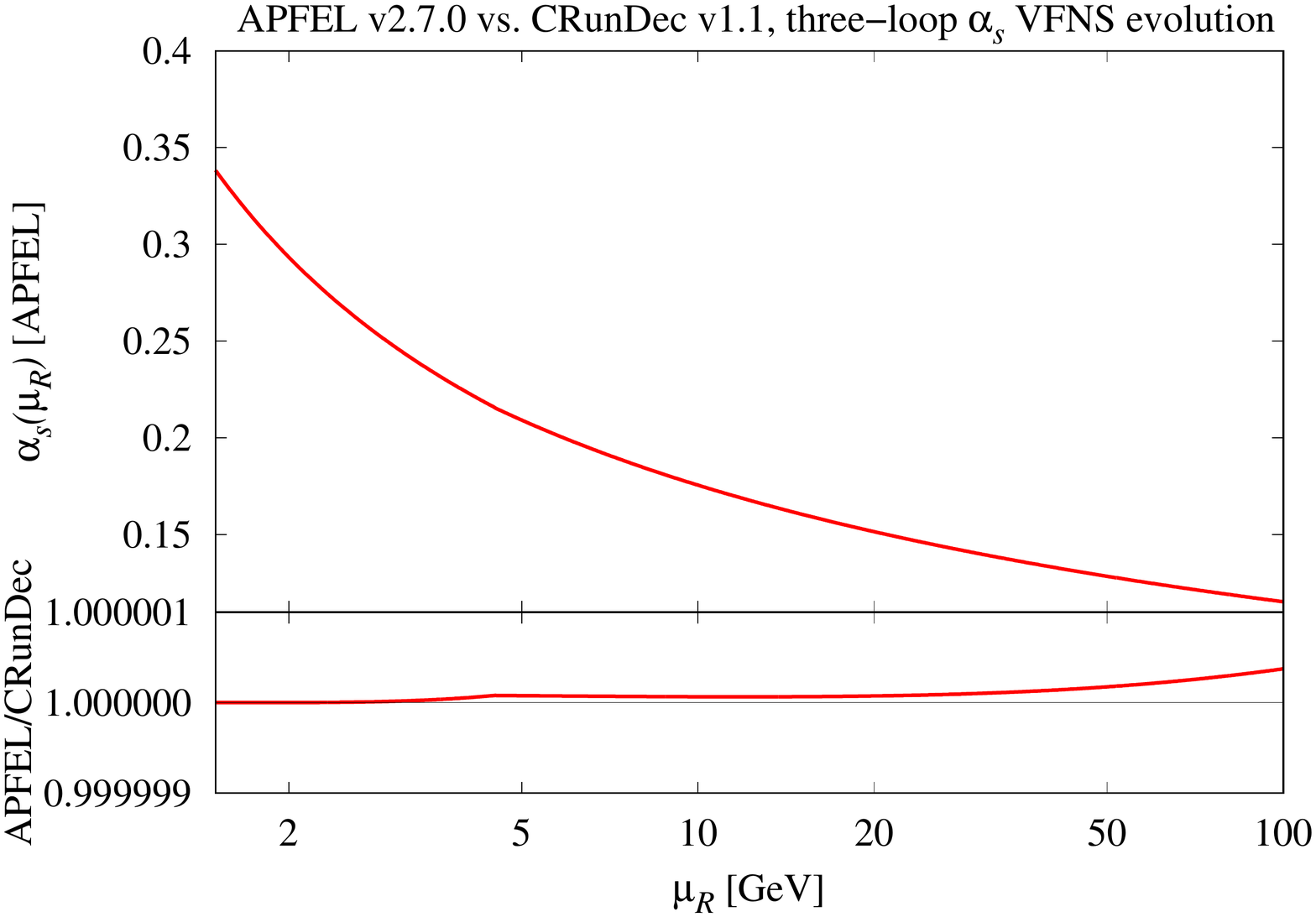}
    \includegraphics[width=0.49\textwidth]{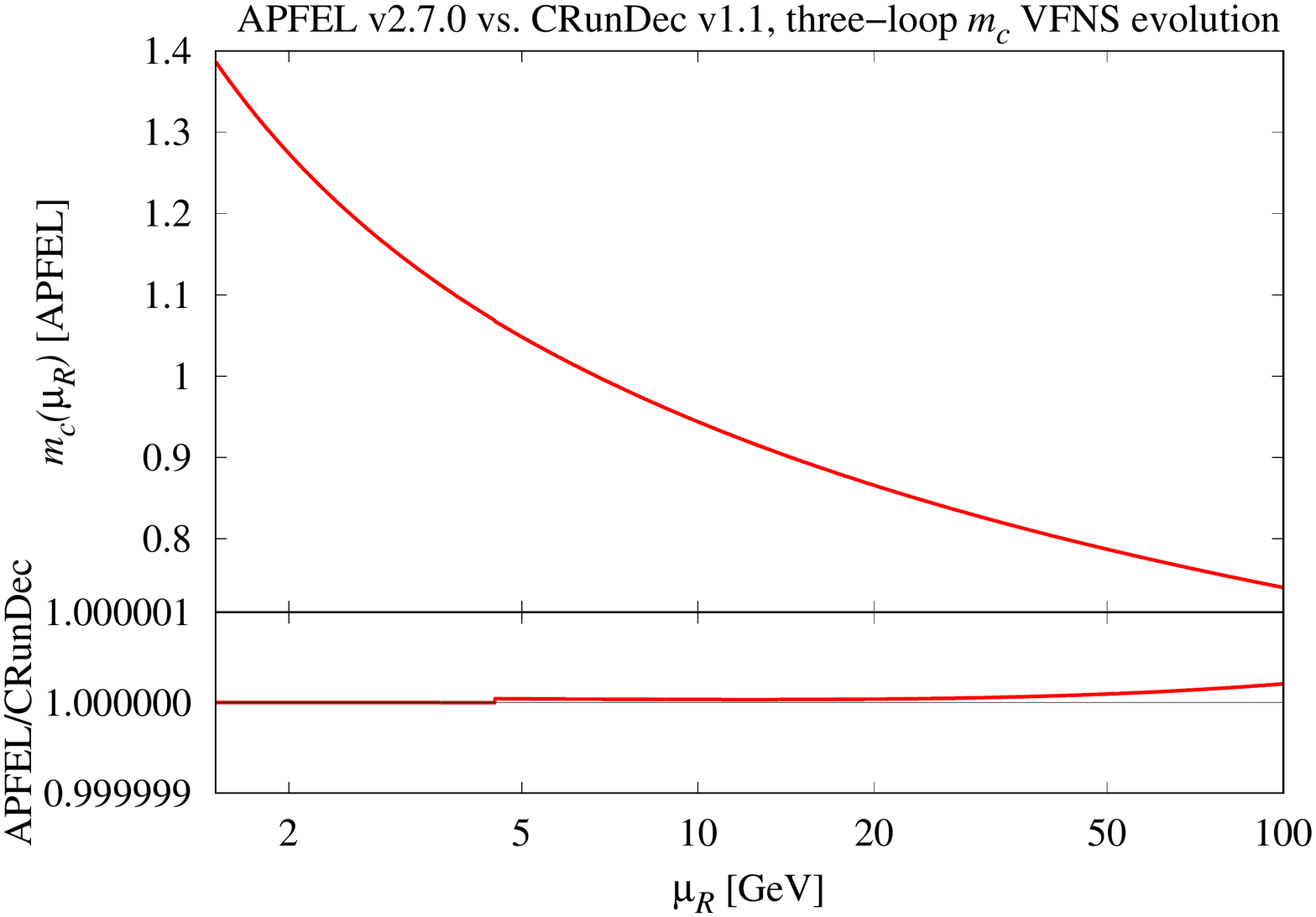}
  \end{center}
  \caption{\label{fig:APFELvsRunDec} Comparison between {\tt APFEL}
    v2.7.0 and {\tt CRunDec} v1.1 for the VFNS RG three-loop evolution
    with $\overline{\rm MS}$ heavy-quark thresholds of the strong
    coupling $\alpha_s$ (left plots) and the $\overline{\rm MS}$ charm
    mass $m_c$ (right plot). The evolution settings are:
    $\alpha_s^{(n_f=3)}(\sqrt{2}\mbox{ GeV}) = 0.35$,
    $m_c^{(n_f=4)}(m_c) = \sqrt{2}\mbox{ GeV}$, and
    $m_b^{(n_f=5)}(m_b) = 4.5\mbox{ GeV}$. The upper insets show the
    strong coupling $\alpha_s$ (left) and the charm mass $m_c$ (right)
    as functions of the renormalization scale $\mu_R$ as returned by
    {\tt APFEL}. In the lower insets the ratios to {\tt CRunDec} are
    displayed showing a relative difference well below $10^{-6}$ over
    the complete range considered.}
\end{figure}

Although the benchmark of the DGLAP evolution already provides an
indirect check of the evolution of $\alpha_s$, we have also performed
a direct check of the VFNS evolution with $\overline{\rm MS}$
heavy-quark thresholds of $\alpha_s$ along with the evolution of the
$\overline{\rm MS}$ charm mass. To this end, we have used the {\tt
  CRunDec} code~\cite{Schmidt:2012az}, which is the {\tt C++} version
of the {\tt Mathematica} package {\tt
  RunDec}~\cite{Chetyrkin:2000yt}. In Fig.~\ref{fig:APFELvsRunDec} we
show the comparison between {\tt APFEL} and {\tt CRunDec} for the
three-loop evolution (NNLO) of the strong coupling $\alpha_s$ (left
plot) and the charm mass $m_c$ (right plot). As is clear from the
lower insets, the agreement between the two codes is excellent. Also
the one- and two-loop evolutions have been checked finding the same
level of agreement.

\begin{figure}
  \begin{center}
    \includegraphics[width=0.49\textwidth]{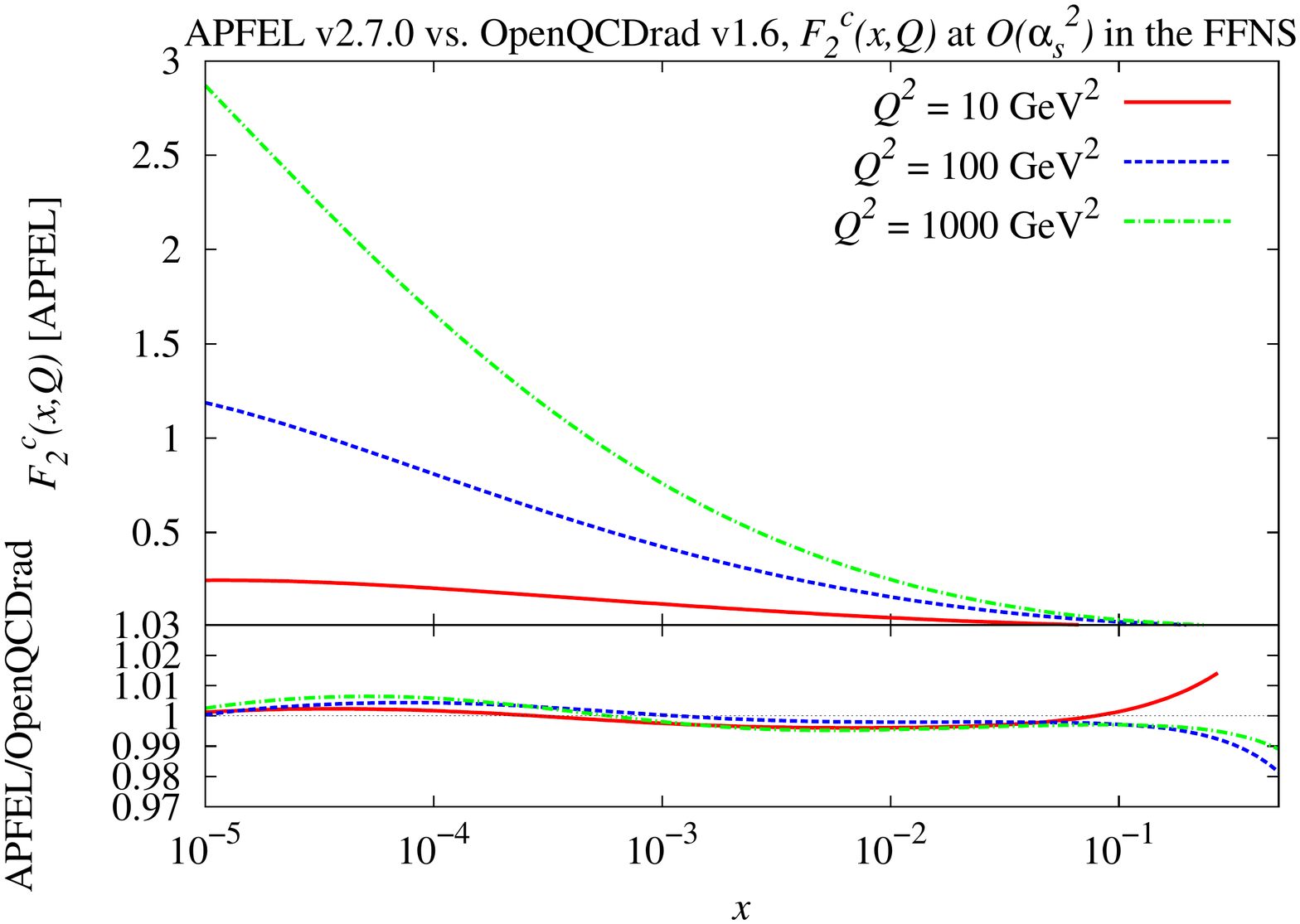}
    \includegraphics[width=0.49\textwidth]{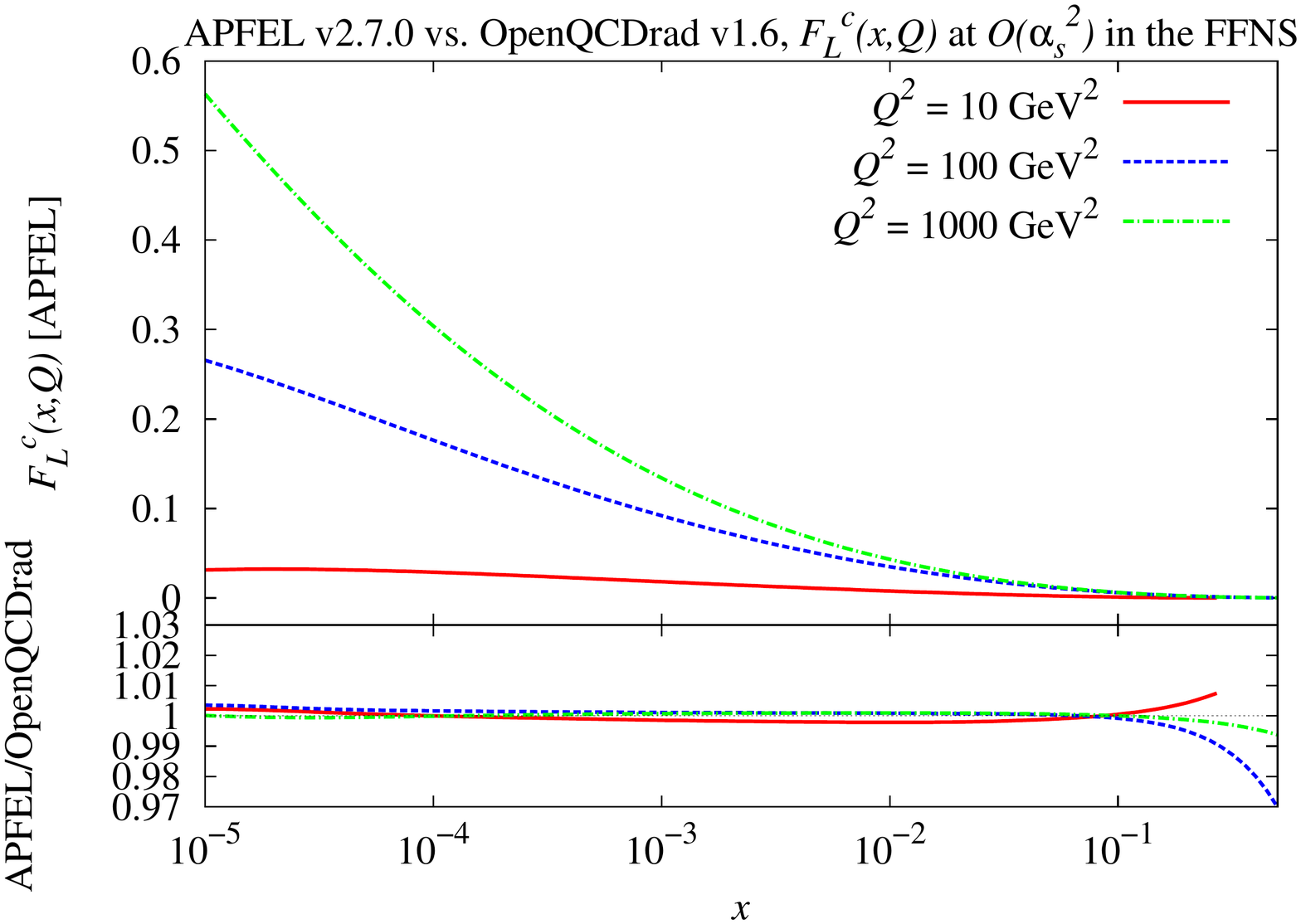}
  \end{center}
  \caption{\label{fig:APFELvsOpenQCDrad} Comparison between {\tt
      APFEL} v2.7.0 and {\tt OPENQCDRAD} v1.6 for the neutral-currents
    massive charm structure functions with $\overline{\rm MS}$
    heavy-quark masses at $\mathcal{O}(\alpha_s^2)$. As an input PDF
    set we have used {\tt MSTW2008nlo68cl\_nf3}~\cite{Martin:2010db}
    from which also the numerical values of $\alpha_s$ and $m_c$ are
    taken. The upper insets show $F_2^c$ (left) and $F_L^c$ (right) as
    functions of $x$ for $Q^2 = 10, 100, 1000$ GeV$^2$ as returned by
    {\tt APFEL}. In the lower insets the ratios to {\tt OPENQCDRAD}
    are displayed showing a relative difference at the per-mil level
    except in the very large-$x$ region where, due to the smallness of
    the predictions, the relative differences tend to increase but
    maintain a good level of absolute accuracy.}
\end{figure}

Finally, we benchmarked the implementation of massive DIS structure
functions ($i.e.$ $F^{(3)}$ in eq.~\ref{eq:FONLLdef}) with
$\overline{\rm MS}$ masses against the public code {\tt OPENQCDRAD}
v1.6~{\cite{openqcdrad}}. {\tt OPENQCDRAD} implements DIS structure
functions in terms of the $\overline{\rm MS}$ heavy-quark masses
following the formalism discussed in
Ref.~\cite{Alekhin:2010sv}. However, as already mentioned above, such
a procedure does not directly correspond to what is needed for the
implementation of the FONLL scheme. In order to make the comparison
with {\tt OPENQCDRAD} possible, we have implemented in {\tt APFEL} a
variant of the FONLL scheme with $\overline{\rm MS}$ masses where, as
done in {\tt OPENQCDRAD} , the RG running of the heavy-quark masses is
expanded and truncated to the appropriate order. In
Fig.~\ref{fig:APFELvsOpenQCDrad} we show the comparison between {\tt
  APFEL} and {\tt OPENQCDRAD} for the exclusive charm neutral-current
structure functions $F_2^c$ (left plot) and $F_L^c$ (right plot) at
$\mathcal{O}(\alpha_s^2)$ for three different values of $Q^2$ and over
a wide range of $x$. As is clear from the lower ratio plots, the
agreement is typically at the per-mil level except in the very
large-$x$ region where, due to the smallness of the predictions, the
relative difference tends to increase but maintains a good level of
absolute accuracy.

To conclude this section, we observe that, referring to
eq.~(\ref{eq:FONLLdef}), the introduction of the $\overline{\rm MS}$
masses does not affect the four-flavor structure function
$F^{(4)}$. The structure function $F^{(3,0)}$ is instead affected by
the transition from pole to $\overline{\rm MS}$ masses. Since we are
not aware of any public code that computes such structure functions, a
direct bechmark has not been possible. However, as a sanity check we
have checked that $F^{(3,0)}$ and $F^{(3)}$ for large values of $Q^2$
tend to the same value, as the definition of $F^{(3,0)}$ requires.

\section{QCD fit settings} \label{sec:fitsettings}

The QCD fits were performed to the combined H1 and ZEUS charm
production cross-section measurements~\cite{Abramowicz:1900rp}
together with the combined HERA1+2 H1 and ZEUS inclusive DIS
cross-section data~\cite{Abramowicz:2015mha}, accounting 
for all given sources of systematic uncertainties.

The kinematic region covered by HERA is constrained by the invariant
mass of the hadronic system of $W > 15$ GeV and the Bjorken scaling
variable of $x < 0.65$, therefore target mass corrections are expected
to have negligible effects and are not discussed in this paper. The
settings of the QCD fits in {\tt xFitter} closely follow those used
for the HERAPDF2.0 PDF extraction~\cite{Abramowicz:2015mha}, with a
few differences related to the specifics of the current analysis which
are motivated in the following.

The nominal result is extracted using the FONLL-C variant of the FONLL
scheme discussed in Sect.~\ref{sec:fonll}. It should be pointed out
that, while being accurate at NNLO for the inclusive DIS cross
sections, the sensitivity to mass corrections of the FONLL-C scheme is
actually NLO. The reason is that at $\mathcal{O}(\alpha_s^0)$ the
FONLL scheme reduces to the parton model which is insensitive to
heavy-quark mass effects. Therefore, the first mass-sensitive term is
$\mathcal{O}(\alpha_s)$ which is the accuracy of the FONLL-A scheme
which would thus provide a LO determination of the charm mass. Both
the FONLL-B and the FONLL-C schemes, instead, include the
$\mathcal{O}(\alpha_s^2)$ massive corrections and thus would both
produce determinations of the mass of the charm accurate at NLO.  The
advantage of FONLL-C with respect to FONLL-B is that it is accurate at
$\mathcal{O}(\alpha_s^2)$ also in the massless sector and thus it is
supposed to provide a better description of the data. In other words,
FONLL-C is the most accurate variant of the FONLL scheme presently
available and as such it will be employed for our determination of
$m_c(m_c)$.

The result obtained in the FONLL scheme is accompanied by an analogous
determination of $m_c(m_c)$ obtained using the FFN scheme with
$\overline{\rm MS}$ masses~\cite{Alekhin:2013nda} at NLO. Access to
the structure functions calculated with the FFN scheme is possible via
the {\tt xFitter} interface to the {\tt OPENQCDRAD}
program~\cite{openqcdrad} using the {\tt QCDNUM} program for the PDF
evolution~\cite{qcdnum}.

The procedure to determine the $\overline{\rm MS}$ charm mass follows
closely the methodology described in Ref.~\cite{Abramowicz:1900rp}. It
involves a series of fits in each of which a set of PDFs is determined
corresponding to numerical values of charm mass ranging between
$m_c(m_c)=1.15$ GeV and $m_c(m_c)=1.60$ GeV with steps of $0.05$ GeV.
For each value of $m_c(m_c)$ a value of global $\chi^2$ is obtained.
The best fit value of $m_c(m_c)$ is determined from the minimum of the
parabolic fit to the resulting $\chi^2$ distribution
and the associated 1-$\sigma$ uncertainty, which reflects the
sensitivity of the data set to the charm mass, is determined as the
$\Delta\chi^2=1$ variation around the minimum.
 
We now discuss the settings of the nominal fits and the variations
that we performed to assess the different sources of uncertainty
deriving from: the PDF parametrization, the model parameters, and the
theoretical assumptions.

The assumption that heavy-quark PDFs are dynamically generated via
gluon splitting at the respective thresholds requires that the
starting scale $Q_0$ at which PDFs are parametrized is below the charm
threshold, which in turn is identified with $m_c(m_c)$. Given the
range in which the scan of $m_c(m_c)$ is done (from 1.1 to 1.6 GeV),
we have chosen to set $Q_0 = 1$ GeV to allow all fits to be
parametrized at the same starting scale. The combinations and the
relative functional forms of the initial scale PDFs have been chosen
following the parametrization scan procedure as performed for the
HERAPDF2.0 determination~\cite{Abramowicz:2015mha}, and the optimal
configuration has been found to be:
\begin{equation}\label{eq:param}
\begin{array}{rclcl}
xg(x) & & &=& A_gx^{B_g}(1-x)^{C_g}-A'_gx^{B'_g}(1-x)^{25},\\
xu_v(x) &=& xu(x)-x\overline{u}(x) &=& A_{u_v}x^{B_{u_v}}(1-x)^{C_{u_v}}(1+E_{u_v}x^2),\\
xd_v(x) &=& xd(x)-x\overline{d}(x) &=& A_{d_v}x^{B_{d_v}}(1-x)^{C_{d_v}},\\
x\bar{U}(x) &=& x\overline{u}(x) &=& A_{\bar{U}}x^{B_{\bar{U}}}(1-x)^{C_{\bar{U}}}(1+D_{\bar{U}}x),\\
x\bar{D}(x) &=& x\overline{d}(x)+x\overline{s}(x) &=& A_{\bar{D}}x^{B_{\bar{D}}}(1-x)^{C_{\bar{D}}}.
\end{array}
\end{equation}
There are $14$ free parameters, since additional constraints were
applied as follows.  The QCD sum rules are imposed at the starting
scale and constrain the normalisation parameters $A_g$, $A_{u_v}$,
$A_{d_v}$.  The light-sea quark parameters that affect the low-$x$
kinematic region $B_{\bar{U}}$ and $B_{\bar{D}}$, as well as the
normalisation parameters $A_{\bar{U}}$ and $A_{\bar{D}}$, are
constrained by the requirement that $\bar{u}\rightarrow \bar{d}$ as $x
\to 0$, leading to the following constraints:
\begin{eqnarray}
  B_{\bar{U}}  &=& B_{\bar{D}} ,\\
  A_{\bar{U}}  &=& A_{\bar{D}}(1-f_s),
\label{eq:constr}
\end{eqnarray}
with $f_s$ being the strangeness fraction of $\bar{D}$ assumed at the
starting scale, $i.e.$ $f_s=\bar{s} / \bar{D}$, because HERA data alone
are not able to provide a precise light-sea flavor separation. The
strangeness fraction for the nominal fits is set to $f_s = 0.4$, as in
the HERAPDF2.0 analysis~\cite{Abramowicz:2015mha}.

In order to estimate the uncertainty associated to the PDF
parametrization, we have considered the following variations with
respect to the nominal configuration:
\begin{itemize}
\item we have moved up the initial scale $Q_0$ from 1 to $\sqrt{1.5}$
  GeV. In the FONLL scheme, this restricted the $m_c(m_c)$ range in
  which we did the scan because we could not use values of the charm
  mass such that $m_c(m_c) < \sqrt{1.5}$ GeV. We were however able to
  perform the parabolic fit in order to find the best fit value of
  $m_c(m_c)$. This complication does not arise in the FFN scheme in
  which there is no threshold crossing.
\item In the $xu_v$ distribution we have included an additional linear
  term so that the last factor in second line of eq.~(\ref{eq:param})
  reads $(1+D_{u_v} x + E_{u_v} x^2)$. After trying different
  variations of the parametrization, we found that this particular
  choice leads to the largest differences.
\end{itemize}

Moving to the model parameters, the values of the bottom and top quark
masses for the nominal fits are chosen to be equal to the PDG values,
defined in the $\overline{\rm MS}$ scheme, $i.e.$ $m_b(m_b)=4.18$ GeV
and $m_t(m_t)=160$ GeV~\cite{Agashe:2014kda}. The value of the strong
coupling is set to $\alpha_s(M_Z) = 0.118$. It should be pointed out
that this value of $\alpha_s$ assumes 5 active flavors. For the FFN
scheme fits, though, one needs to use the value of $\alpha_s$ with 3
active flavors. In order to find this value one has to evolve
$\alpha_s(M_Z)$ down to below $m_c(m_c)$ in the VFN scheme and evolve
back to $M_Z$ with 3 active flavors. We have computed the value of
$\alpha_s$ with 3 active flavors for each of the values of $m_c(m_c)$
considered.

The uncertainty associated to model parameters will be estimated by
considering the following variations:
\begin{itemize}
\item the bottom mass has been moved up and down by 0.25 GeV, $i.e.$
  $m_b(m_b) = 3.93$ GeV and $m_b(m_b) = 4.43$ GeV. The magnitude of
  the variation is actually much larger than the present uncertainty
  on the bottom mass and thus our choice is meant to provide a
  conservative estimate of the associated uncertainty.
\item The variation of the strong coupling follows the recent PDF4LHC
  prescription~\cite{Butterworth:2015oua}. In particular, we have
  considered the conservative variation up and down by 0.0015 with
  respect to the nominal value, $i.e.$ $\alpha_s(M_Z) = 0.1165$ and
  $\alpha_s(M_Z) = 0.1195$.
\item Finally, we considered the value of the strangeness fraction
  introduced in eq.~(\ref{eq:constr}) as being a model parameter and
  we have thus varied it up and down by 0.1 around the nominal value
  considering $f_s=0.3$ and $f_s=0.5$.
\end{itemize}

We finally turn to the theory assumptions and their variations. These
mostly concern unknown higher-order corrections and the most common
way to estimate them is by varying the renormalization and the
factorization scales $\mu_R$ and $\mu_F$. As nominal scales in our
analysis we have chosen $\mu_R^2 =\mu_F^2 = Q^2$ for both the
FONLL\footnote{A scale choice involving the heavy-quark mass would
  lead to technical complications with the FONLL matching as
  implemented in {\tt APFEL}. However, we have checked that the more
  commonly used scales $\mu_R^2 =\mu_F^2 = Q^2+4m_c(m_c)^2$ produce a
  very marginal difference in the determination of $m_c(m_c)$ in the
  FFN scheme.} and the FFN scheme analyses. Another possible source of
theoretical uncertainty in the FONLL scheme is the presence of the
damping factor discussed in Sect.~\ref{sec:fonll} which is meant to
suppress unwanted subleading terms and whose explicit form in the
nominal fits is given in eq.~(\ref{eq:dampingfactor}).

The theoretical uncertainty associated to the missing higher-order
corrections has been estimated as follows:
\begin{itemize}
\item the factorization and renormalization scales were varied by a
  factor 2 up and down with respect to the nominal values, that is
  choosing $\mu_R^2 =\mu_F^2 = Q^2 / 2$ and $\mu_R^2 =\mu_F^2 = 2Q^2$.
  Such variations have been applied only to the heavy-quark components
  of the structure functions, while the light part has been left
  unchanged. The reason for this is that, in order to estimate the
  theoretical uncertainty associated to the determination of
  $m_c(m_c)$, we want to perform scale variations only in the part of
  the calculation sensitive to this parameter, which is clearly the
  charm structure function (for consistency, the same variation was
  applied also to the bottom structure functions).
\item As already mentioned, the FONLL damping factor represents a
  further source of uncertainty. It has the role of suppressing
  unwanted subleading terms but the particular way in which this
  suppression is implemented is somewhat arbitrary. To assess the
  impact of our particular choice on the determination of $m_c(m_c)$,
  we have changed the suppression power around the nominal one,
  considering the following functional form:
  \begin{equation}
    D_p(Q,m_c)=\theta(Q^2-m_c^2)\left(1-\frac{m_c^2}{Q^2}\right)^p\,,
  \end{equation}
  with $p=1,4$.
\end{itemize}

In addition, to assure the applicability of perturbative QCD and to
keep higher-twist corrections under control, a cut on $Q^2$ is imposed
on the fitted data. Our nominal cut is $Q^2 > Q_{\rm min}^2=3.5$
GeV$^2$. The choice of the value of $Q_{\rm min}^2$ requires some
care; an extensive discussion on the impact of varying it on the
determination of $m_c(m_c)$ is given in
Sect.~\ref{sec:Q2minDependence}.

To conclude this section, we observe that the self-consistency of the
input data set and the good control of the systematic uncertainties
enable the determination of the experimental uncertainties in the PDF
fits using the tolerance criterion of $\Delta \chi^2=1$.

\section{Results}\label{sec:results}

In this section we will present the result for our the determination
of the value $m_c(m_c)$ in the $\overline{\mbox{MS}}$ renormalization
scheme using the FONLL scheme with its associated set of
uncertainties.

\begin{figure*}
  \begin{center}
    \includegraphics[angle=270,width=0.9\textwidth]{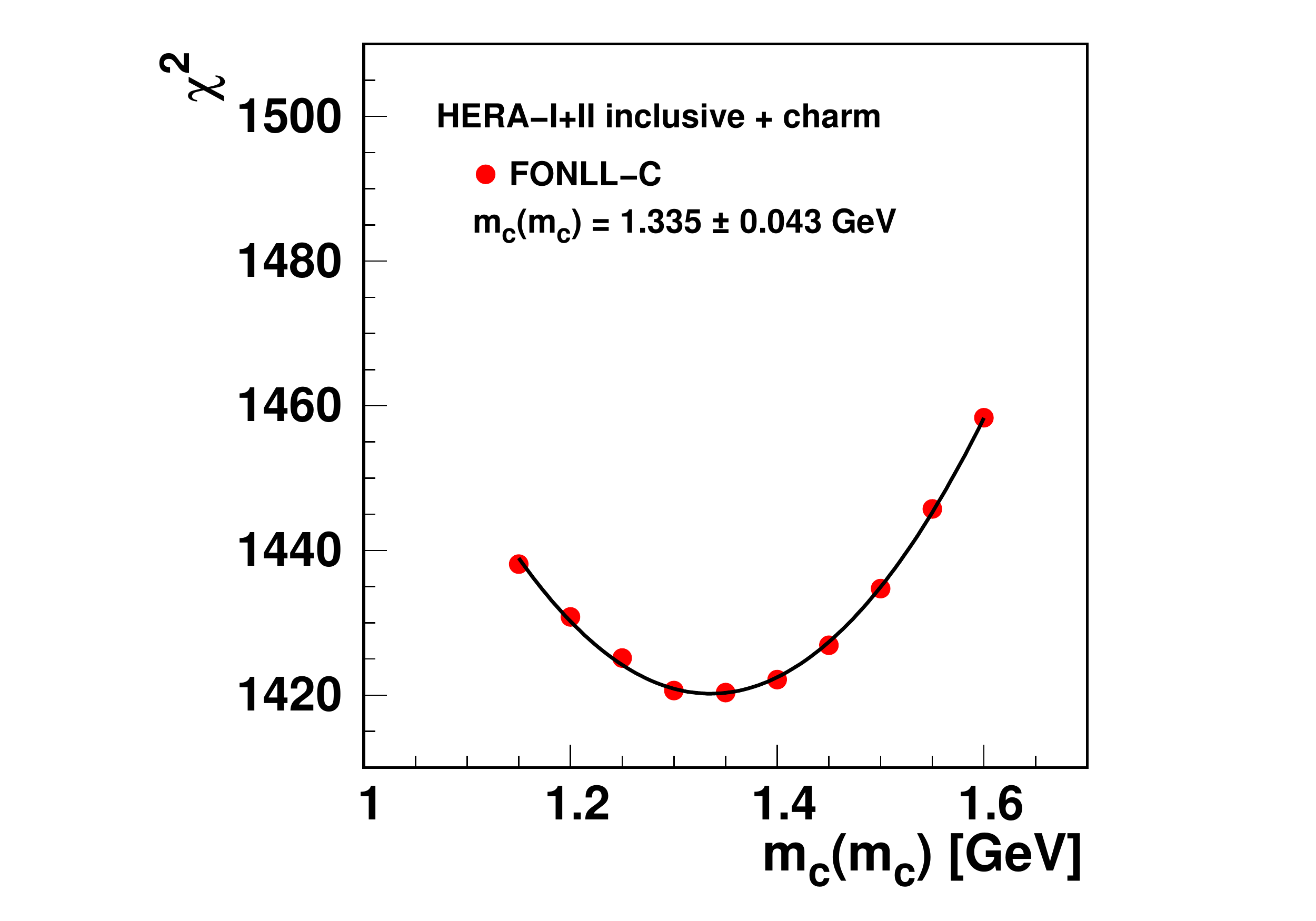}
  \end{center}
  \caption{\label{fig:fonllC} Parabolic fit to the global $\chi^2$ as
    a function of $m_c(m_c)$ in the FONLL-C scheme with nominal
    settings.}
\end{figure*}

\begin{figure*}
  \begin{center}
    \includegraphics[angle=270,width=0.45\textwidth]{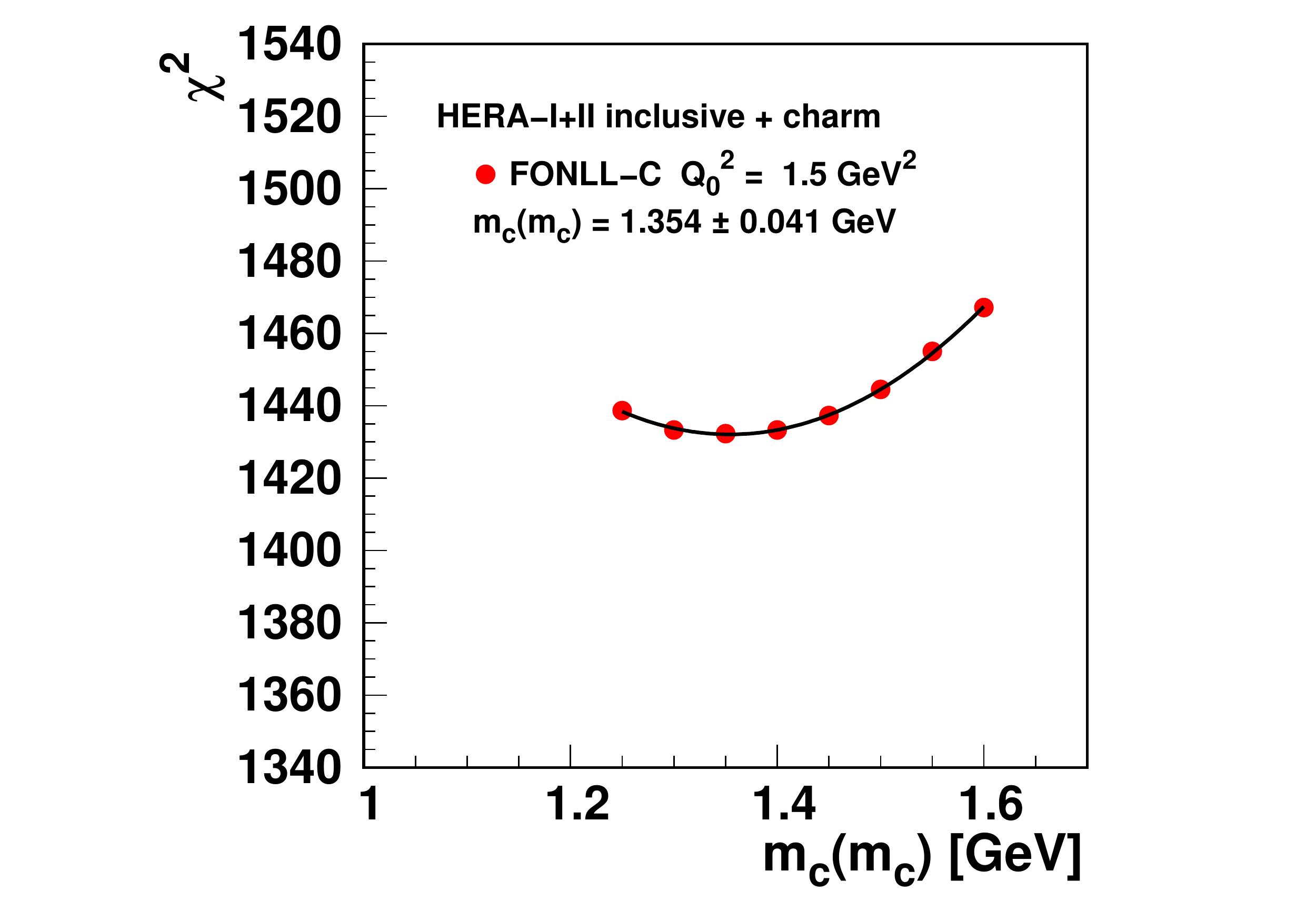}\includegraphics[angle=270,width=0.45\textwidth]{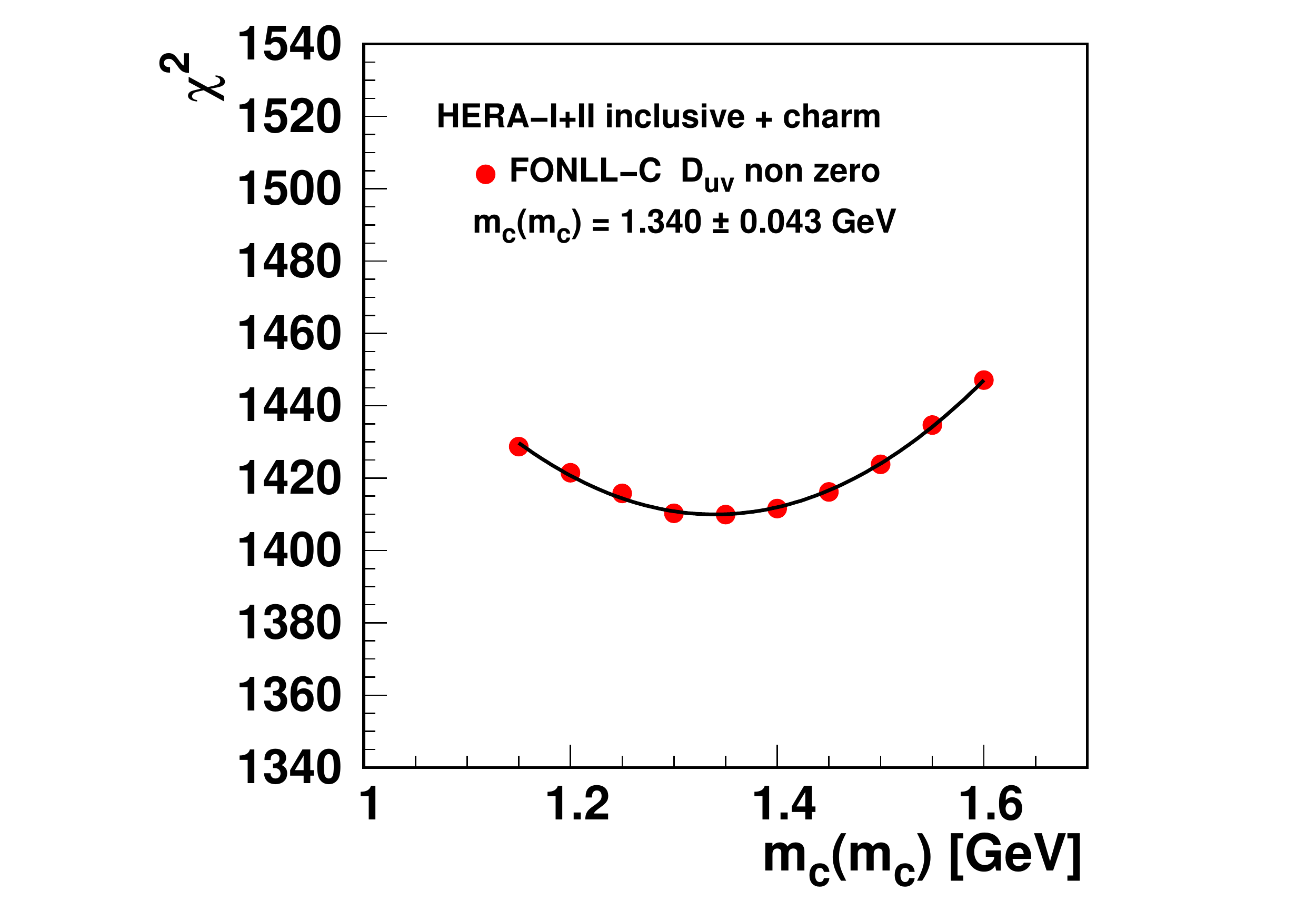}\\
    \includegraphics[angle=270,width=0.45\textwidth]{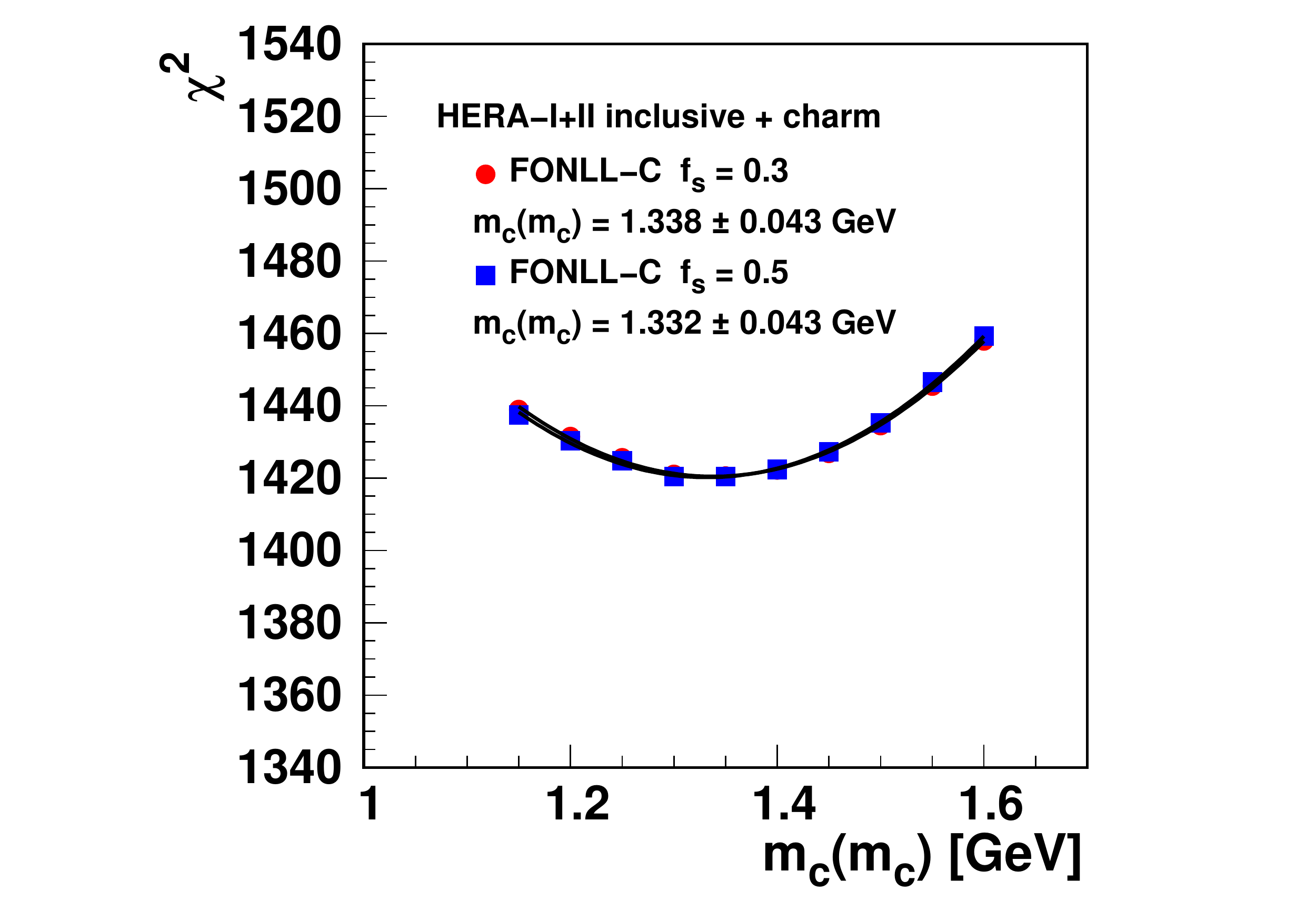}\includegraphics[angle=270,width=0.45\textwidth]{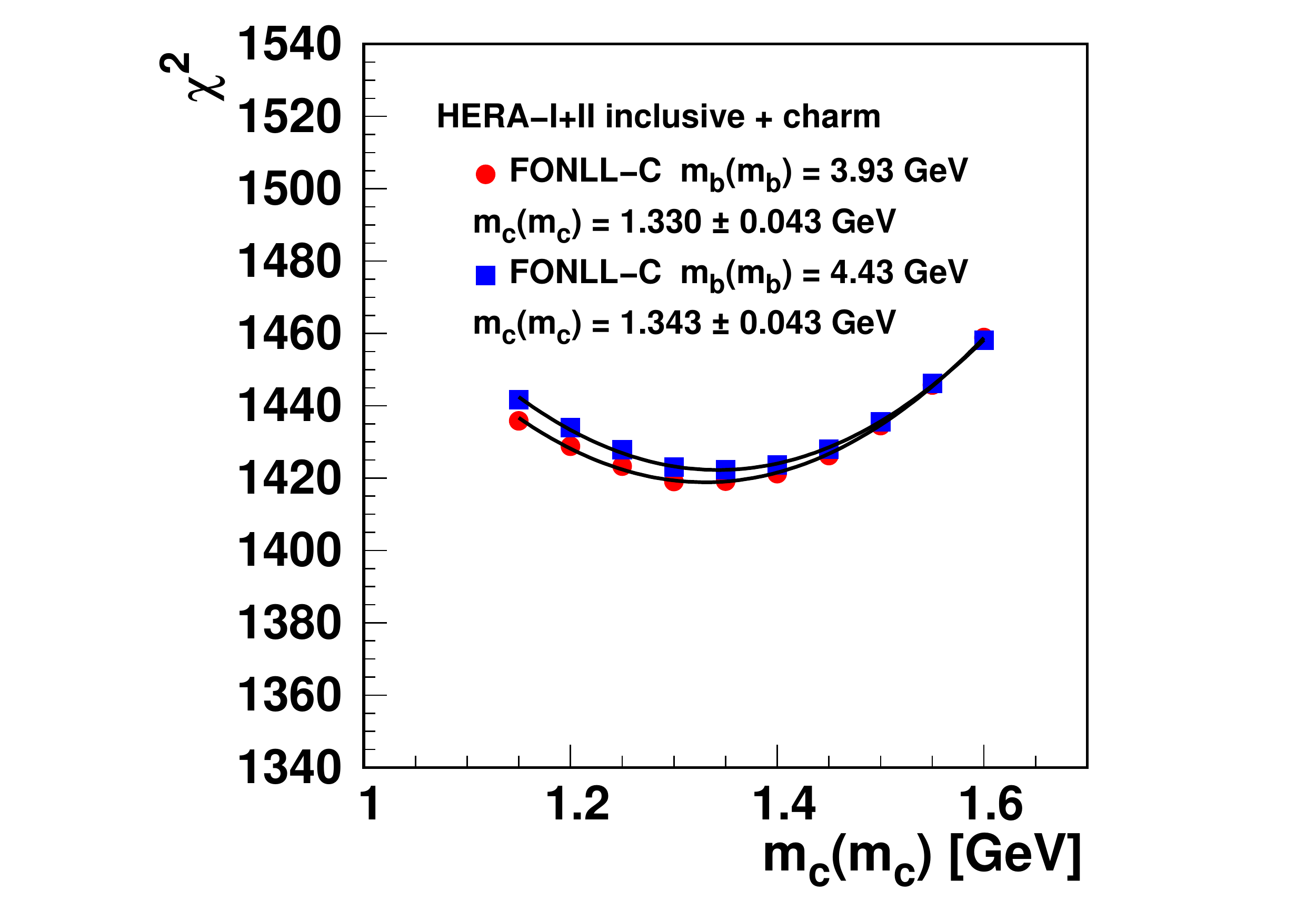}\\
    \includegraphics[angle=270,width=0.45\textwidth]{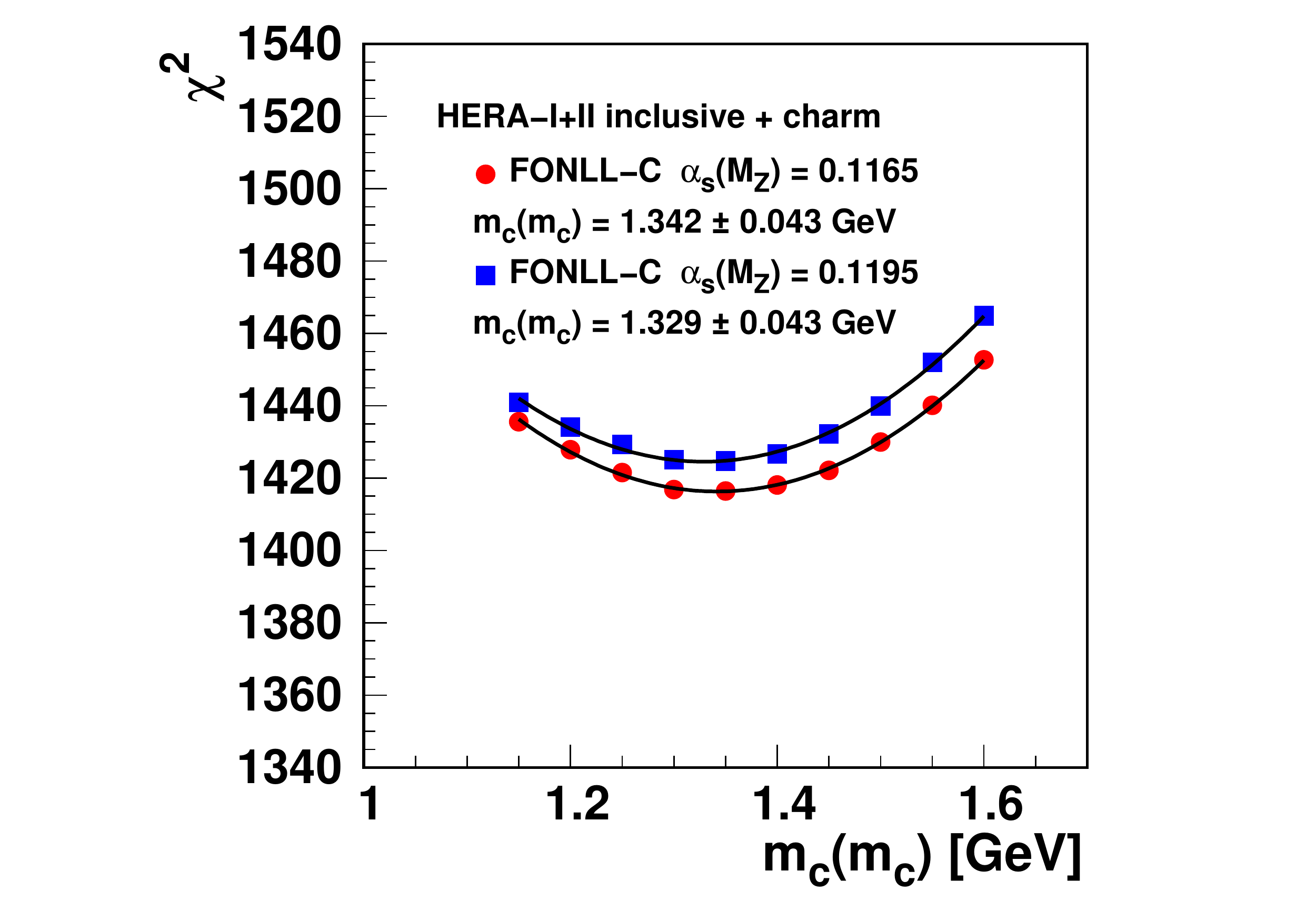}\includegraphics[angle=270,width=0.45\textwidth]{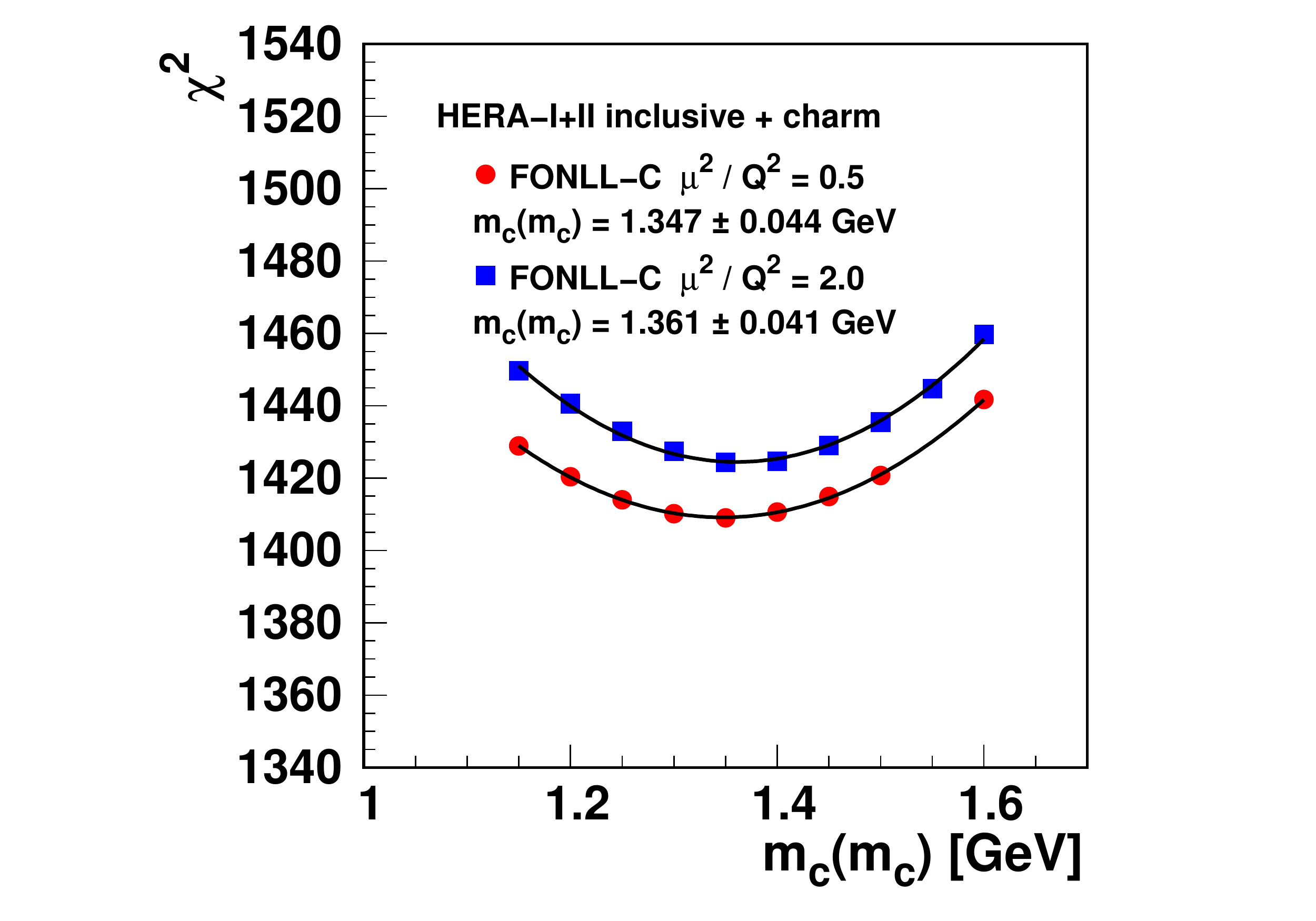}\\
    \includegraphics[angle=270,width=0.45\textwidth]{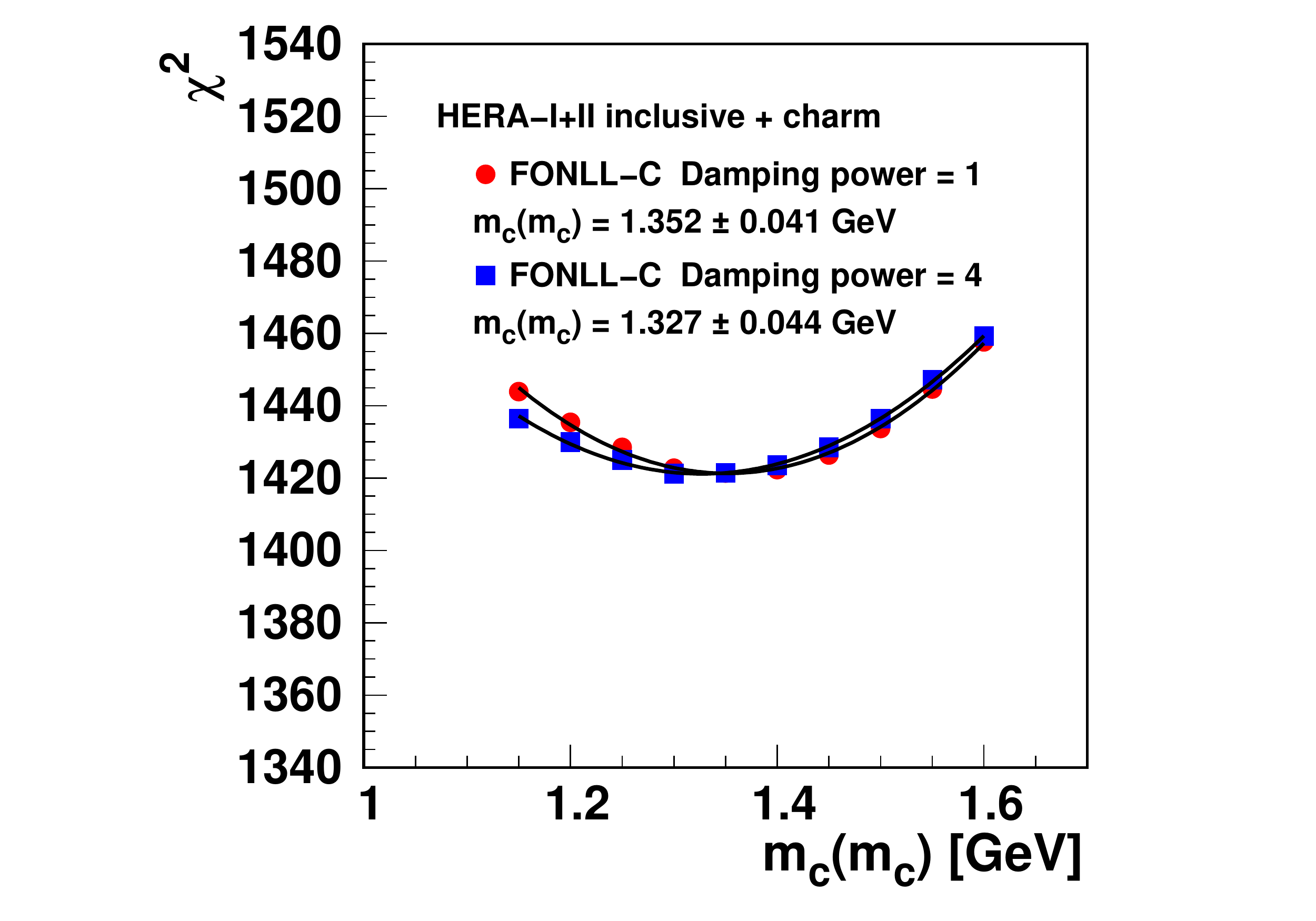}
  \end{center}
  \caption{\label{fig:variations} Parabolic fits to the global
    $\chi^2$'s as functions of $m_c(m_c)$ in the FONLL-C scheme for
    all variations performed to estimate the non-experimental
    uncertainties on $m_c(m_c)$.}
\end{figure*}

The parabolic fit to the global $\chi^2$ as a function of $m_c(m_c)$
is shown in Fig.~\ref{fig:fonllC} and yields a best fit value and its
1-$\sigma$ experimental uncertainty equal to
$m_c(m_c)=1.335 \pm 0.043$ GeV.  An estimate of the parametric, model,
and the theoretical uncertainties, performed following the procedure
described in Sect.~\ref{sec:fitsettings}, is summarised in the second
column of Tab.~\ref{tab:model} and leads to our final result:
\begin{equation}\label{eq:fonll}
m_c(m_c)=1.335 \pm 0.043\mbox{(exp)}^{+0.019}_{-0.000}\mbox{(param)}^{+0.011}_{-0.008}\mbox{(mod)}^{+0.033}_{-0.008}\mbox{(th)} \mbox{ GeV.} 
\end{equation}
An illustration of the deviations, again determined through parabolic
fits, caused by the variations employed to determine the parametric,
model, and theoretical uncertainties is given in
Fig.~\ref{fig:variations}.
\begin{table*}
  \centering
  \begin{tabular*}{\textwidth}{@{\extracolsep{\fill}}lcc@{}}
    \hline
    variation                          & FONLL-C  & FFN \\
    \hline
    \hline
    central                             &  $1.335 \pm 0.043$ &  $1.318\pm 0.054$\\
    \hline
    \hline
    $Q_0^2=1.5$                   & $1.354\;[+0.019]$ & $1.329\;[+0.011]$ \\
    \hline 
    $D_{uv}$ non-zero            & $1.340\;[+0.005]$ & $1.308\;[-0.010]$ \\
    \hline 
    \hline 
    $f_s=0.3$                        &   $1.338\;[+0.003]$ & $1.320\;[+0.002]$  \\
    $f_s=0.5$                        &   $1.332\;[-0.003]$ & $1.315\;[-0.003]$ \\
    \hline
    $m_b(m_b)=3.93$ GeV     & $1.330\;[-0.005]$ & $1.312\;[-0.006]$  \\
    $m_b(m_b)=4.43$ GeV     & $1.343\;[+0.008]$ & $1.324\;[+0.006]$ \\
    \hline
    $\alpha_s(M_Z)=0.1165$ & $1.342\;[+0.007]$ & $1.332\;[+0.014]$ \\
    $\alpha_s(M_Z)=0.1195$ & $1.329\;[-0.006]$ & $1.300\;[-0.018]$\\
    \hline
    \hline
    $\mu_F^2=\mu_R^2=2\cdot Q^2$ & $1.347\;[+0.012]$ & $1.314\;[-0.004]$ \\
    $\mu_F^2=\mu_R^2= Q^2/2$        & $1.361\;[+0.026]$ & $1.363\;[+0.045]$ \\
    \hline
    FONLL Damping power = 1 & $1.352\;[+0.017]$ & -- \\
    FONLL Damping power = 4 & $1.327\;[-0.008]$ & -- \\
    \hline
    \end{tabular*}
    \caption{\label{tab:model} List of the variations performed to
      estimate the non-experimental uncertainties on $m_c(m_c)$ with
      the respective results obtained in the FONLL-C scheme and in the FFN scheme at NLO.}
\end{table*}

After we have determined the best fit value of the charm mass in
eq.~(\ref{eq:fonll}), we have used the central value to perform a
further fit in the FONLL-C scheme (nominal fit). In
Tab.~\ref{tab:chi2} we report the partial $\chi^2$'s over the number
of data points for each subset along with the total correlated
$\chi^2$, the logarithmic penalty, and the total $\chi^2$ per degree
of freedom.
\begin{table}
  \begin{center}
    \begin{tabular}{lp{2.57cm}}
      \hline
  Data Set     & $\chi^2$ \\ 
  \hline
  Charm cross section H1-ZEUS combined & 44 / 47  \\ 
  HERA1+2 CCep & 43 / 39  \\ 
  HERA1+2 CCem & 55 / 42  \\ 
  HERA1+2 NCem & 218 / 159  \\ 
  HERA1+2 NCep 820 & 67 / 70  \\ 
  HERA1+2 NCep 920 & 439 / 377  \\ 
  HERA1+2 NCep 460 & 220 / 204  \\ 
  HERA1+2 NCep 575 & 219 / 254  \\ 
  Correlated $\chi^2$  & 104  \\ 
  Log penalty $\chi^2$  & +12  \\ 
  \hline
  Total $\chi^2$ / d.o.f.  & 1420 / 1178  \\ 
  \hline
    \end{tabular}
  \end{center}
  \caption{\label{tab:chi2} $\chi^2$'s resulting from the fit in the
    FONLL-C scheme using the best fit value of the charm mass
    $m_c(m_c) = 1.335$ GeV. The partial $\chi^2$'s per data point
    along with the total correlated $\chi^2$, the logarithmic penalty,
    and the total $\chi^2$ / d.o.f. are reported, as defined in Ref.~\cite{Aaron:2012qi}.}
\end{table}

\begin{figure*}
  \begin{center}
    \includegraphics[angle=0,width=0.48\textwidth]{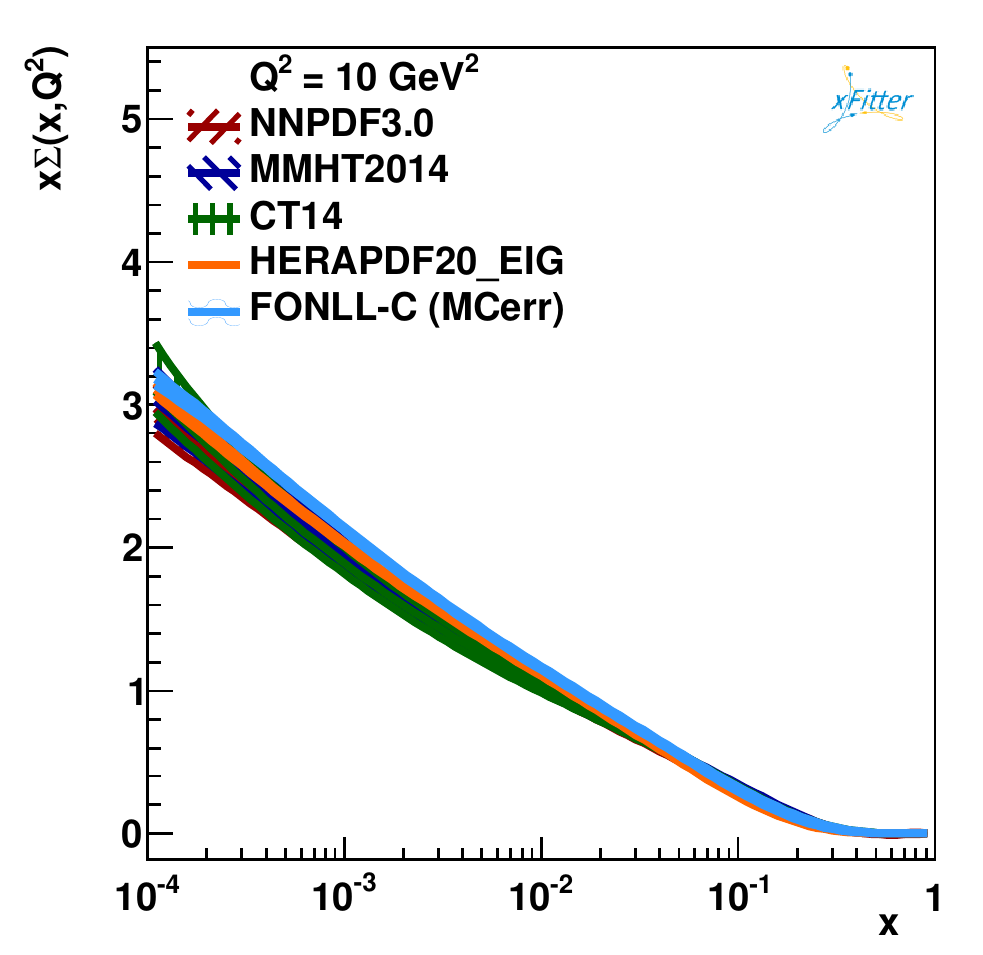}\includegraphics[angle=0,width=0.48\textwidth]{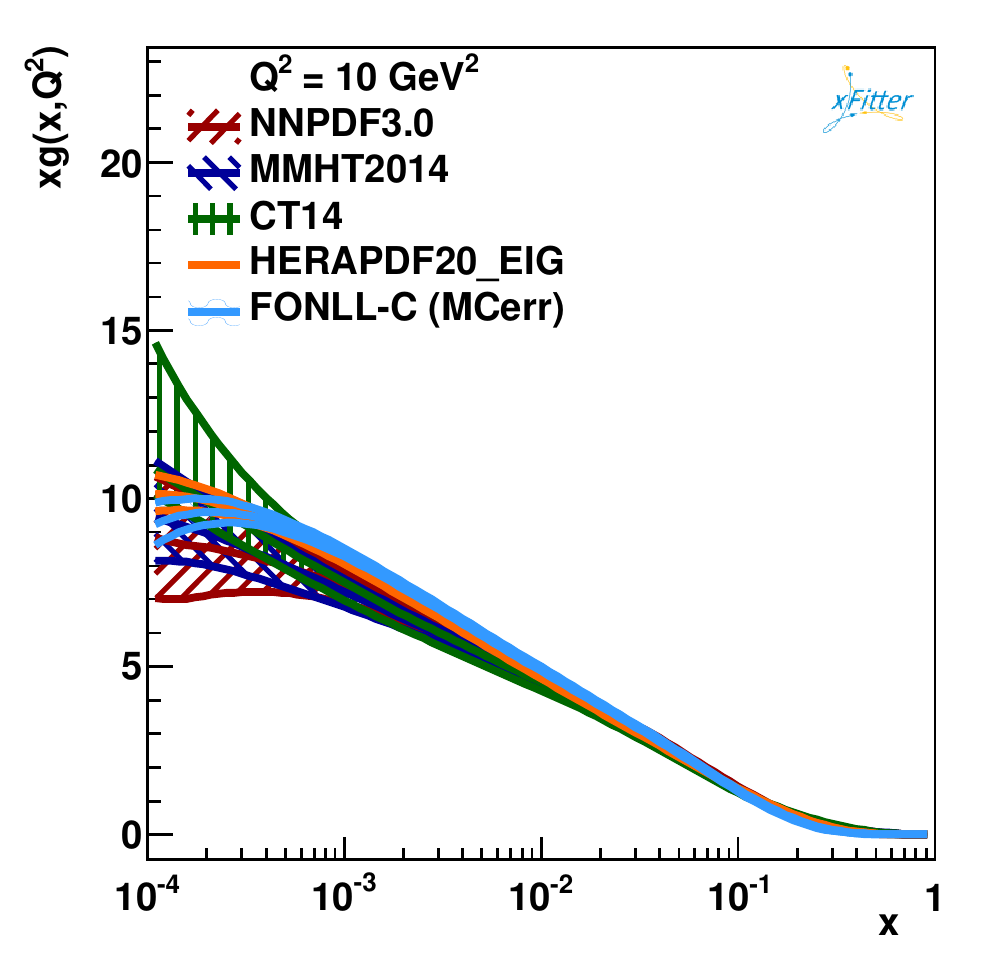}
  \end{center}
  \caption{\label{fig:pdfs} Comparison at $Q^2 = 10$ GeV$^2$ of the
    singlet (left plot) and gluon (right plot) distributions from the
    nominal FONLL-C fit with other PDF sets determined using GM-VFN
    schemes: HERAPDF2.0, CT14, MMHT14, NNPDF3.0.}
\end{figure*}

As an illustration, the singlet and the gluon PDFs extracted from the
nominal fits are compared with other GM-VFNS PDF sets:
CT14\cite{Dulat:2015mca}, HERAPDF2.0\cite{Abramowicz:2015mha},
MMHT14\cite{Thorne:2015caa}, NNPDF3.0\cite{Ball:2014uwa}. They are
shown in Fig.~\ref{fig:pdfs} at the scale $Q^2=10$ GeV$^2$, where the 
the experimental uncertainties from the nominal fits on PDFs are estimated using Monte Carlo
procedure with the root mean square estimated from 500 replica.  An overall
good agreement is observed.

\begin{figure*}
  \begin{center}
    \includegraphics[width=0.7\textwidth]{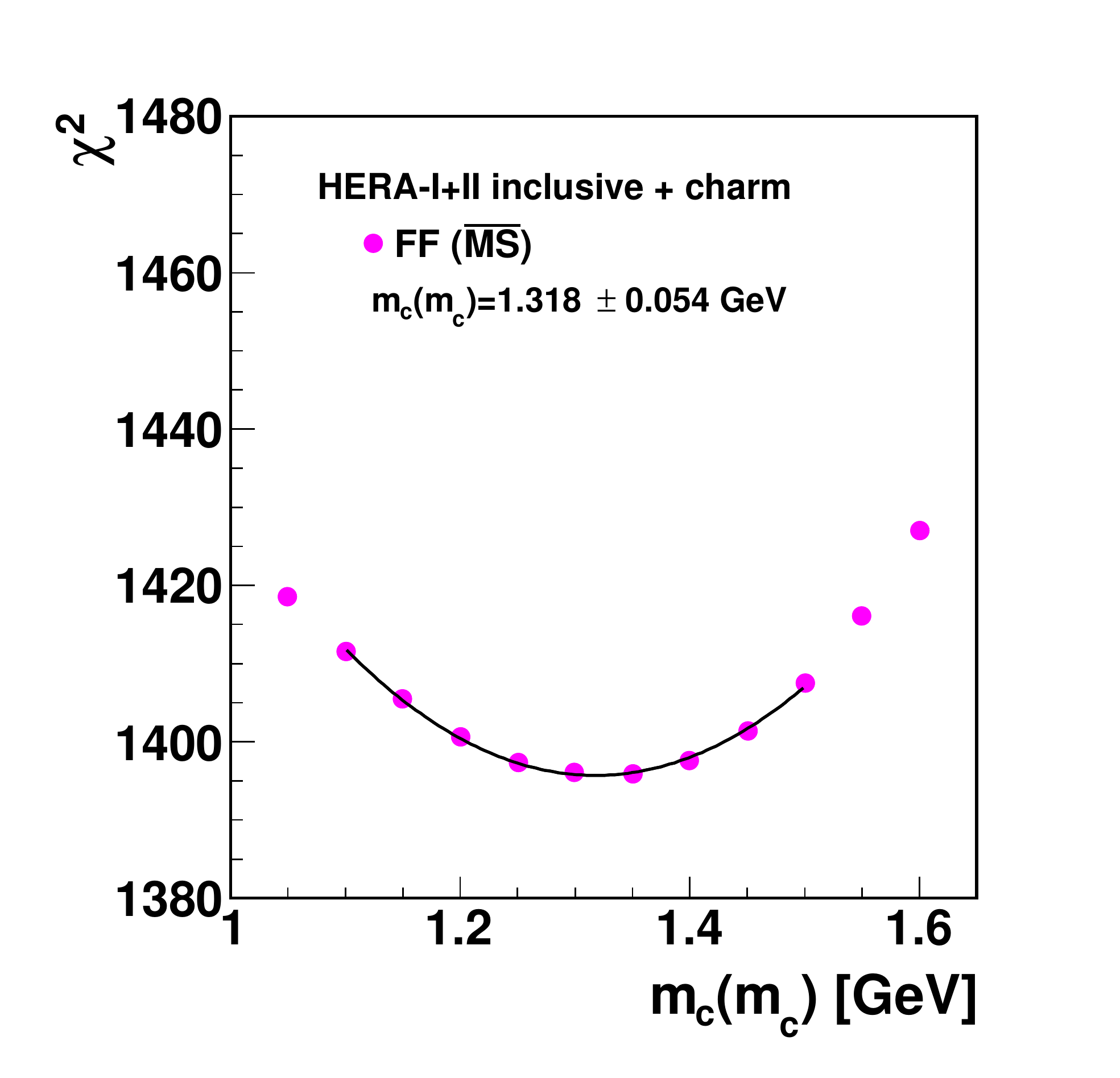}
  \end{center}
  \caption{\label{fig:ffABM} Parabolic fit to the global $\chi^2$ as
    a function of $m_c(m_c)$ in the FFN scheme at NLO with nominal
    settings.}
\end{figure*}

The FONLL determination of $m_c(m_c)$ presented above is supported by
an analogous determination in the FFN scheme at NLO. The corresponding
parabolic fit with the associated experimental uncertainty is shown in
Fig.~\ref{fig:ffABM}. Also in this case a full characterization of the
non-experimental uncertainty has beed achieved by carrying out the
same parametric, model, and theory variations (except for the
variation of the damping factor which is specific of the FONLL
scheme). The results of the variation in the FFN scheme are reported
in the third column of Tab.~\ref{tab:model}. The final result is:
\begin{equation}\label{eq:ffab}
m_c(m_c) = 1.318\pm 0.054\mbox{(exp)}^{+0.011}_{-0.010}\mbox{(param)}^{+0.015}_{-0.019}\mbox{(mod)}^{+0.045}_{-0.004}\mbox{(th)}\mbox{ GeV}\,,
\end{equation}
which is in agreement with the FONLL determination given in
eq.~(\ref{eq:fonll}).

It is interesting to notice that we observe a reduced scale dependence
in the FONLL scheme as compared to the FFN scheme. We ascribe this
effect to the fact that the leading contributions in the FONLL scheme
involve both gluon- and quark-initiated processes; typically the
contributions from gluon processes decrease with the scale, while the
contributions from quark processes tend to increase. Conversely, the
FFN scheme is mostly driven by gluon processes the contributions of
which (along with $\alpha_s$) tend to be monotonic in $\mu$ leading to
larger scale variations\footnote{We thank Fred Olness for this
  interesting observation.}.

As discussed Sect.~\ref{sec:MatchingConditions}, the running of the
$\overline{\mbox{MS}}$ heavy-quark masses in the VFN scheme, exactly
like the running of $\alpha_s$ and PDFs, is not univocally defined at
the heavy-quark thresholds due to the presence of the so-called
matching conditions. In particular, when giving the value of the mass
at one of the heavy-quark thresholds, one should also specify whether
this corresponds to the value immediately below or above the threshold
itself. This is typically done by complementing the value with the
number of active flavors used in the computation. In fact, in general
$m_c^{(N_f=3)}(m_c) \neq m_c^{(N_f=4)}(m_c)$. On theoretical grounds,
this difference is relevant when comparing a determination obtained in
a VFN scheme like FONLL with a determination obtained in the ($N_f=3$)
FFN scheme: in the latter one automatically determines
$m_c^{(N_f=3)}(m_c)$, while in the former it is more natural to
extract $m_c^{(N_f=4)}(m_c)$. However, eqs.~(\ref{eq:MarchMhUp})
and~(\ref{eq:MarchMhDown}) tell us how the two values are connected up
to $\mathcal{O}(\alpha_s^2)$ and applying eq.~(\ref{eq:MarchMhUp}) to
the central value eq.~(\ref{eq:fonll}) one gets
$m_c^{(N_f=3)}(m_c) = 1.339$ GeV, that is a difference of 0.004 GeV as
compared to the nominal value which is well within the current
uncertainty on $m_c(m_c)$.  We can then conclude that, even though
providing a value $m_c(m_c)$ is ambiguous if the number of active
flavors is not specified, the magnitude of the ambiguity is currently
not large enough to significantly affect the current determinations.

\subsection{Comparison to other results}\label{sec:comparison}

It is interesting to compare our results with the past determinations
of $\overline{\mbox{MS}}$ charm mass $m_c(m_c)$ using a similar
methodology (also see Ref.~\cite{Behnke:2015qja,Alekhin:2013qqa,Gao:2013wwa} 
for previous comparisons).

The analysis of Ref.~\cite{Alekhin:2012vu} was performed in the ABM11
framework~\cite{Alekhin:2012ig} using the FFN scheme at NLO and at
approximate NNLO and based on world data for DIS from HERA, and
fixed-target DIS experiments and Tevatron Drell-Yan data. While the
analysis in Ref.~\cite{Alekhin:2012vu} was performed including the
same exclusive charm cross-section data used in this study, it did not
include the HERA1+2 combined inclusive cross-section data set which
was not available at the time, but used instead the HERA combined data
from run 1 only. An earlier analysis~\cite{Alekhin:2012un} used a
partial charm dataset only, with correspondingly larger uncertainties,
while a subsequent analysis~\cite{Alekhin:2013qqa} investigated the
correlation between the measurement of $m_c(m_c)$ and the strong
coupling constant.

The analysis of Ref.~\cite{Gao:2013wwa} is instead based on the
CT10NNLO global analysis, and uses the S-ACOT-$\chi$ GM-VFN scheme
discussed, $e.g.$, in Ref.~\cite{Guzzi:2011ew}. It is based on a
slightly wider data set as it includes LHC jet production data and
also a set of older $F_2^c$ measurements at HERA~\cite{Adloff:2001zj}
that are not included in the more recent combined charm data. The
authors of Ref.~\cite{Gao:2013wwa} provide a set of four
determinations deriving from different strategies to convert the
pole-mass definition into $\overline{\mbox{MS}}$. They also provide a
separate estimate of the uncertainty due to the
$\mathcal{O}(\alpha_s^3)$ corrections for one of the four strategies
essentially by varying the parameter that governs a generalized
version of the rescaling variable $\chi$.

Finally, a determination of the charm mass $m_c(m_c)$ was produced by
the H1 and ZEUS collaborations in the framework of the HERAPDF QCD
analysis in the same publication in which the charm cross-section
measurements employed in our study were
presented~\cite{Abramowicz:1900rp}. That determination also used only
the HERA combined inclusive data from run 1~\cite{Aaron:2009aa}.

\begin{figure*}
  \begin{center}
    \includegraphics[angle=180,width=0.5\textwidth]{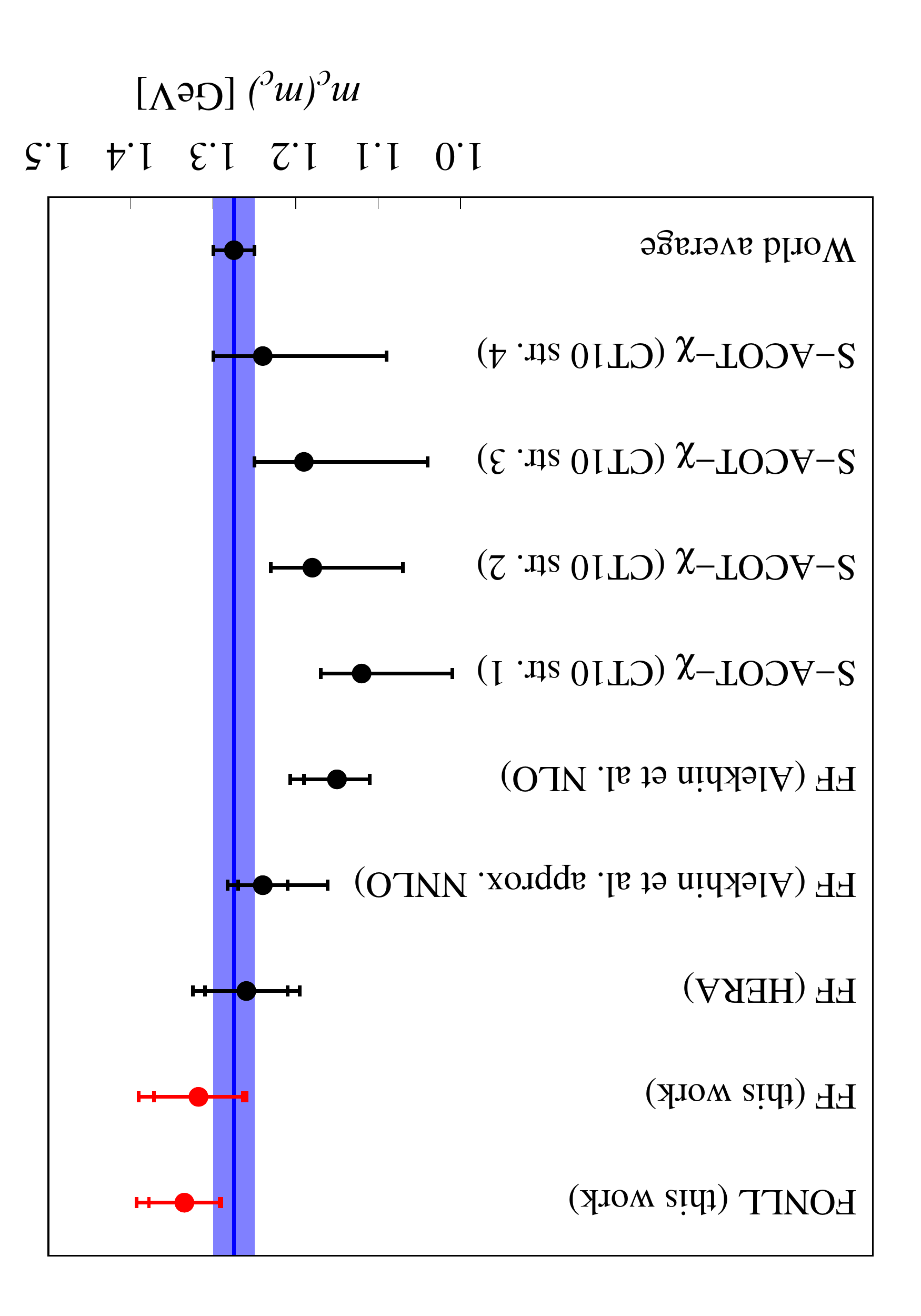}
  \end{center}
  \caption{\label{fig:McComparison} Graphical representation of the
    determinations reported in Tab.~\ref{tab:results}. The inner error
    bars display the experimental uncertainty while the outer error
    bars (when present) are obtained as a sum in quadrature of all
    uncertainty sources. The blue vertical band represents the world
    average and provides a reference for all other determinations.}
\end{figure*}

In Tab.~\ref{tab:results} we report the numerical values for the
$m_c(m_c)$ determinations listed above along with our results
and the world average value~\cite{PhysRevD.86.010001}. 
A short clarification about the nomenclature of the uncertainties
reported in Tab.~\ref{tab:results} is in order. In
Sect.~\ref{sec:fitsettings} we discussed extensively the meaning of
the uncertainties associated to our determinations. In doing so, we
tried to be consistent with the previous determinations, nevertheless
some differences remain. As far as the determination in
Ref.~\cite{Abramowicz:1900rp} is concerned, while their definition of
``(exp)'' and ``(param)'' essentially coincides with ours, their
``(model)'' uncertainty includes the variation of the cut in $Q^2$
(that we will discuss separately in Sect.~\ref{sec:Q2minDependence})
but does not include the $\alpha_s$ variation, which is instead quoted
separately. In addition, the authors do not quote any scale variation
uncertainty. The nomenclature of Ref.~\cite{Alekhin:2012vu} is also
different from ours. Apart from the common ``(exp)'' uncertainty, for
the NLO determination the authors only quote the ``(scale)''
uncertainty, which essentially coincides with our ``(th)'' (even
though the FONLL ``(th)'' uncertainty also accounts for the variation
of the damping factor), while for the approximate NNLO determination
they also quote a ``(th)'' uncertainty which, differently from our
nomenclature, accounts for the uncertainty on the approximated
expressions used at $\mathcal{O}(\alpha_s^3)$. Finally, the
determinations in Ref.~\cite{Gao:2013wwa} only quote the experimental
uncertainty (the asymmetric uncertainties are due to the use of a
generic second-degree polynomial to fit the $\chi^2$ profiles).
A graphical representation of Tab.~\ref{tab:results} is shown in
Fig.~\ref{fig:McComparison} where the inner error bars display the
experimental uncertainty while the outer error bars (when present) are
obtained as a sum in quadrature of all uncertainty sources. The blue
vertical band represents the world average and provides a reference
for all other determinations. It is clear that, while the spread of
the current determinations of $m_c(m_c)$ from DIS data covers a pretty
large range, they are generally in agreement with the world
average. As far as our determinations in particular are concerned, we
observe that, apart from being consistent with each other and with the
world average, they also present competitive uncertainties. This is
particularly relevant for the FONLL determination because this is the
first time that this scheme is employed for a direct determination of
the charm mass.
\begin{table*}
  \centering
  \begin{tabular*}{\textwidth}{@{\extracolsep{\fill}}ll@{}}
    \hline
    scheme  & $m_c(m_c)$ [GeV] \\
    \hline
    \hline
    FONLL (this work) & $1.335\pm0.043\mbox{(exp)}^{+0.019}_{-0.000}\mbox{(param)}^{+0.011}_{-0.008}\mbox{(mod)}^{+0.033}_{-0.008}\mbox{(th)}$\\
    FFN (this work)     & $1.318\pm0.054\mbox{(exp)}^{+0.011}_{-0.010}\mbox{(param)}^{+0.015}_{-0.019}\mbox{(mod)}^{+0.045}_{-0.004}\mbox{(th)}$\\
    \hline
    FFN (HERA)~\cite{Abramowicz:1900rp} & $1.26 \pm 0.05\mbox{(exp)} \pm 0.03\mbox{(mod)} \pm 0.02\mbox{(param)} \pm 0.02(\alpha_s)$ \\
    \hline
    FFN (Alekhin \textit{et al.})~\cite{Alekhin:2012vu}  & $1.24 \pm 0.03({\rm
                                exp})^{+0.03}_{-0.02}({\rm
                                scale})^{+0.00}_{-0.07}({\rm th})$
                                (approx. NNLO)\\ 
     & $1.15 \pm 0.04({\rm exp})^{+0.04}_{-0.00}({\rm scale})$ (NLO)\\ 
    \hline
    S-ACOT-$\chi$ (CT10)~\cite{Gao:2013wwa}  & 
                                                     $1.12^{+0.05}_{-0.11}$
                                        (strategy 1)\\
    & $1.18^{+0.05}_{-0.11}$ (strategy 2)\\
    & $1.19^{+0.06}_{-0.15}$ (strategy 3)\\
    & $1.24^{+0.06}_{-0.15}$ (strategy 4)\\
    \hline
    \hline
    World average~\cite{PhysRevD.86.010001} & $1.275 \pm 0.025$ \\
    \hline
  \end{tabular*}
  \caption{\label{tab:results} List of the recent determinations of
    $m_c(m_c)$ from fits to DIS data along with the
    determinations extracted in this work. The PDG world average value
    is also reported for reference.}
\end{table*}

Fig.~\ref{fig:McComparison} shows that our determinations tend to be
larger than the world average while most of the previous
determinations place themselves below it.  Detailed investigations
show that the largest contribution to this difference arises from the
use of to the new combined HERA1+2 combined inclusive cross section
measurements that are employed for the first time to determine the
charm mass and that, as we will discuss in
Sect.~\ref{sec:Q2minDependence}, tend to prefer larger values of
$m_c(m_c)$.

\subsection{Cross-checks}\label{sec:crosschecks}

It is worth mentioning that we have also employed the variants A and B
of the FONLL scheme discussed in Sect.~\ref{sec:fonll} to determine
$m_c(m_c)$. While the FONLL-A scheme is accurate to LO in the massive
sector and thus does not produce a reliable determination of the charm
mass, the FONLL-B has the same formal accuracy in the massive sector
as FONLL-C and indeed it leads to a determination comparable to that
given in eq.~(\ref{eq:fonll}) both for the central value and the
uncertainties.  It is interesting to notice that the FONLL-B scheme in
the low-energy region resembles very closely the FFN scheme at NLO.
In particular, both schemes are accurate to $\mathcal{O}(\alpha_s^2)$
in the massive sector and to $\mathcal{O}(\alpha_s)$ in the light
sector. As a matter of fact, we find that the experimental uncertainty
associated to the FONLL-B determination is very close to the FFN one
quoted in eq.~(\ref{eq:ffab}), which in turn is around 20\% larger
than that associated to the FONLL-C determination. This suggests that
the $\mathcal{O}(\alpha_s^2)$ corrections to the light sector that are
present in the FONLL-C scheme, which depend on the heavy-quark mass by
means of diagrams in which a gluon plits into a pair of heavy quarks,
provide a further constraint on $m_c(m_c)$.

Finally, we have also attempted a determination in the FFN scheme
using the approximate NNLO massive structure functions as implemented
in {\tt OPENQCDRAD}. However, we did not pursue a full
characterization of the uncertainties because we believe that this
determination, while giving a quantitative indication of the effect of
the NNLO corrections, cannot claim an NNLO accuracy and thus does not
add anything to our NLO determinations.

\subsection{Discussion on the $Q_{\rm min}^2$ dependence of the mass
  determination}\label{sec:Q2minDependence}

Our determination of $m_c(m_c)$ given in eq.~(\ref{eq:fonll}) was
obtained cutting off all data with $Q^2 < Q_{\rm min}^2 = 3.5$
GeV$^2$. The necessity of such a cut stems from the fact that
low-energy data are hard to describe for two main reasons: the large
value of $\alpha_s$ with consequent large higher-order corrections,
and sizable higher-twist corrections. In addition, as pointed out in
Ref.~\cite{Caola:2009iy}, the low-$Q^2$ region (low-$x$, in fact)
might be affected by deviations from the fixed-order DGLAP evolution
whose description might require small-$x$ perturbative resummation.
The dependence on $Q_{\rm min}^2$ of fits to HERA data has already
been discussed in the context of the inclusive measurements only. In
this section, we will address this issue considering also the HERA
charm production data.

\begin{figure*}
  \begin{center}
    \includegraphics[angle=270,width=0.49\textwidth]{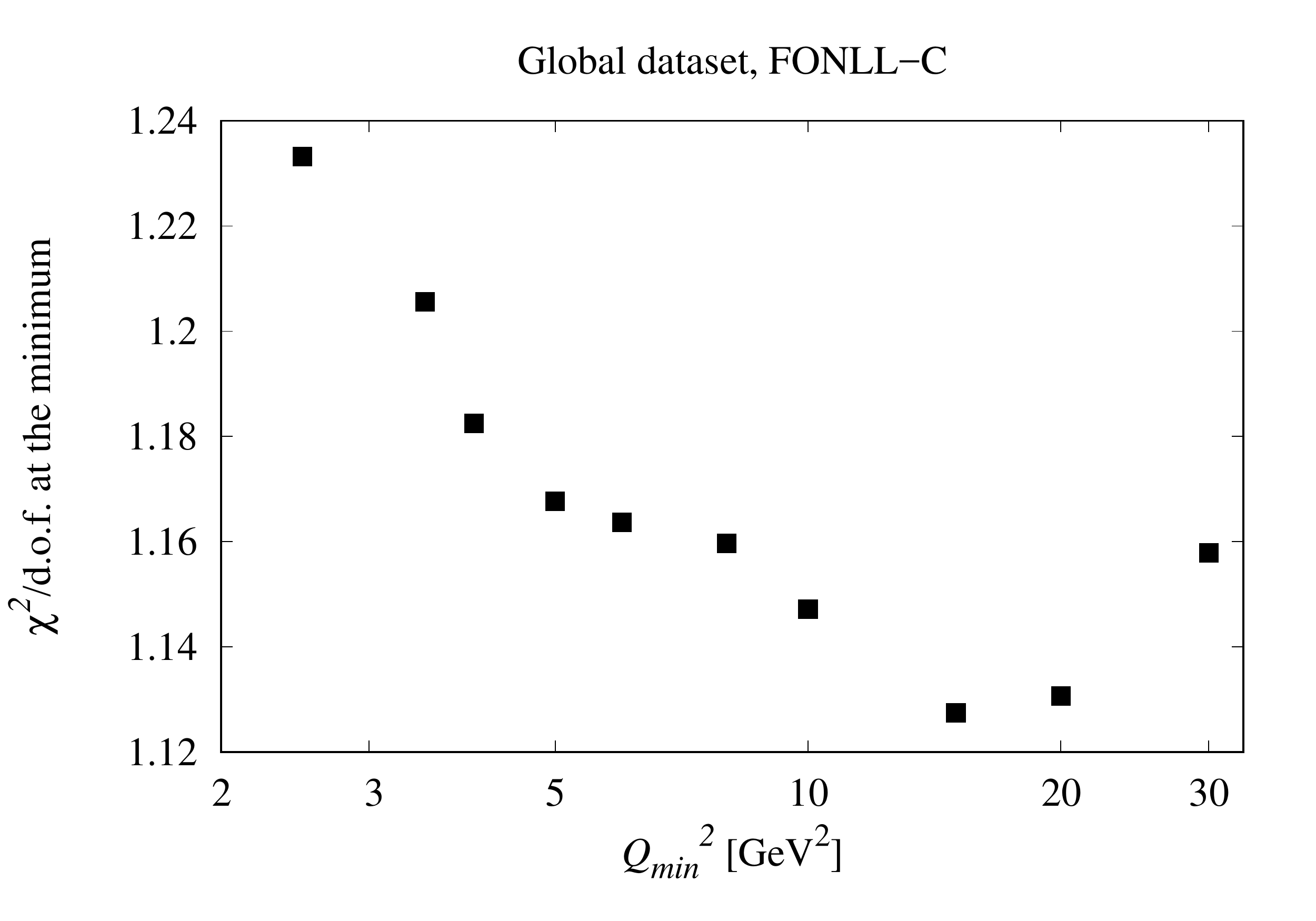}
    \includegraphics[angle=270,width=0.49\textwidth]{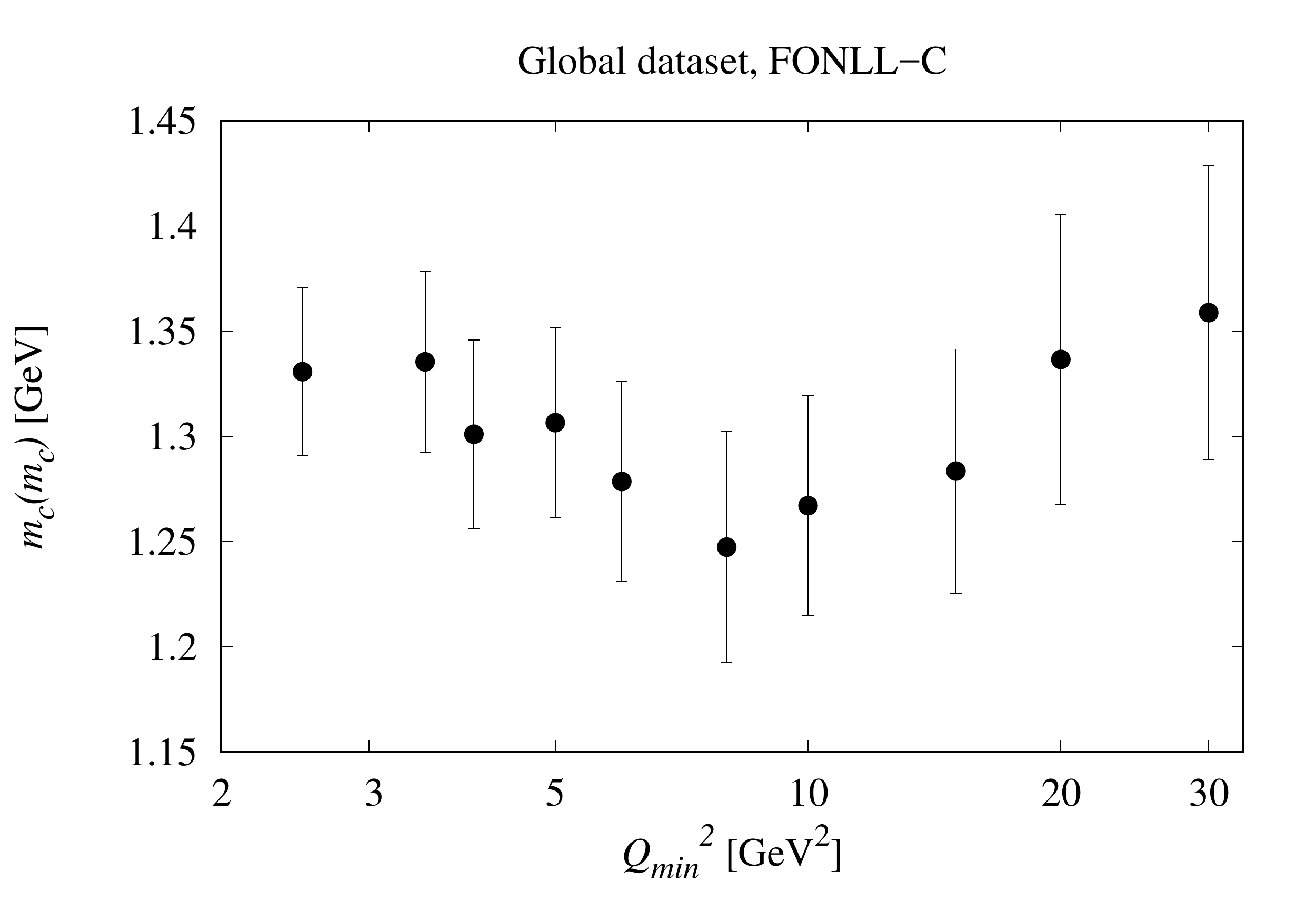}
  \end{center}
  \caption{\label{fig:q2min} Left plot: dependence of the global
    $\chi^2$ / d.o.f. as a function of $Q_{\rm min}^2$. Right plot:
    dependence of the global best fit value of $m_c(m_c)$ with the
    associated experimental uncertainty as a function of
    $Q_{\rm min}^2$. Both plots have been obtained using the FONLL-C
    scheme.}
\end{figure*}

The particular value of $Q_{\rm min}^2$ used in our analysis (3.5
GeV$^2$) was determined by requiring a good fit quality but
maintaining a good sensitivity to $m_c(m_c)$. This is illustrated in
Fig.~\ref{fig:q2min} where the global $\chi^2$ per degree of freedom
is plotted as a function of $Q_{\rm min}^2$ in the left panel while
the best fit of $m_c(m_c)$ is plotted as a function of $Q_{\rm min}^2$
in the right panel.  Looking at the left panel it is clear that, as
expected, the global $\chi^2$ improves as more and more low-energy
data are excluded from the fit. On the other hand, the right plot
shows that the experimental uncertainty associated to $m_c(m_c)$ gets
larger and larger as $Q_{\rm min}^2$ increases indicating that, again
as expected, the sensitivity to $m_c(m_c)$ deteriorates if low-energy
data are excluded. In the light of the plots in Fig.~\ref{fig:q2min},
we conclude that $Q_{\rm min}^2 = 3.5$ GeV$^2$ represents a good
compromise between a good description of the full data set and a good
sensitivity to $m_c(m_c)$.

\begin{figure*}
  \begin{center}
    \includegraphics[angle=270,width=0.49\textwidth]{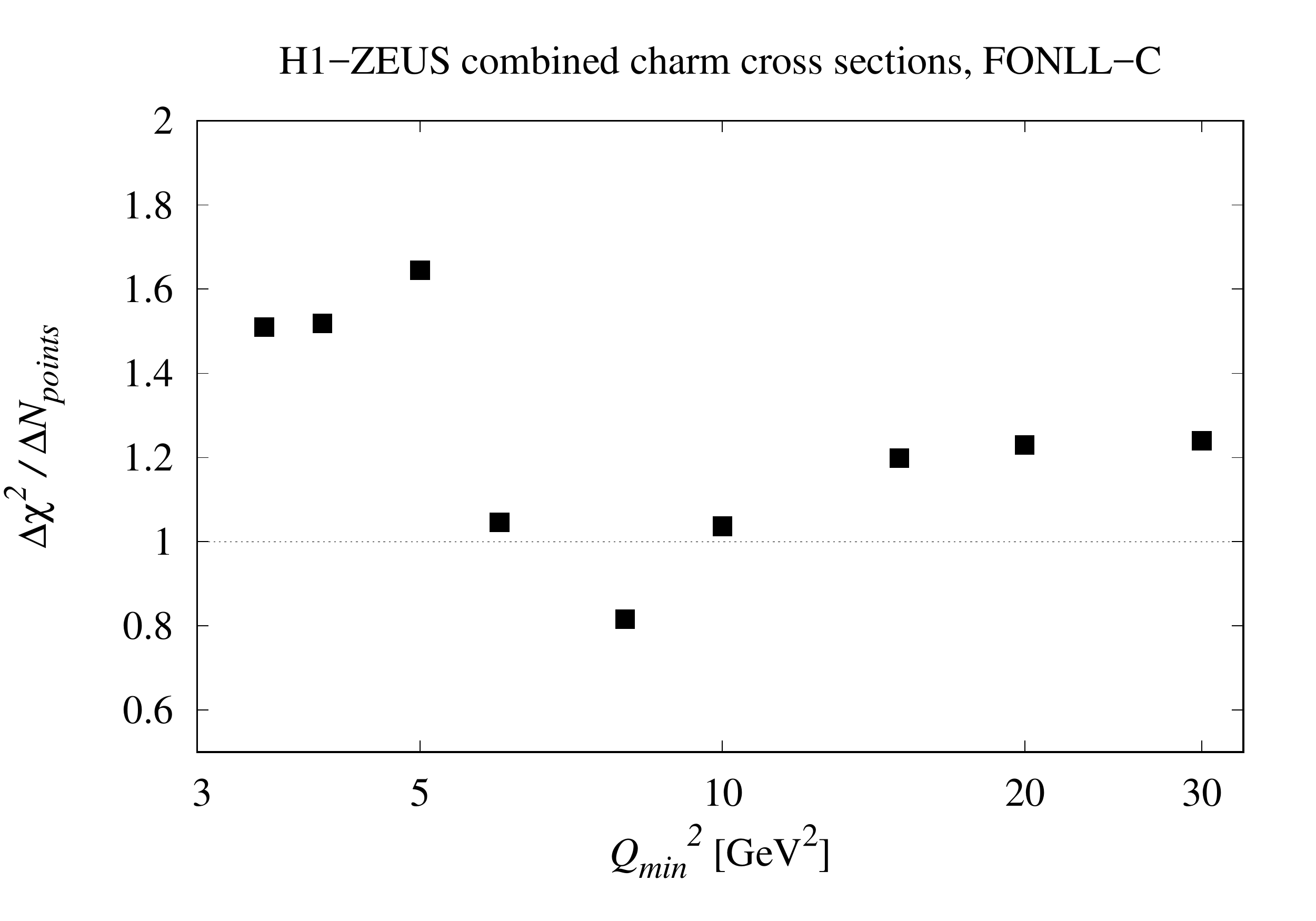}
    \includegraphics[angle=270,width=0.49\textwidth]{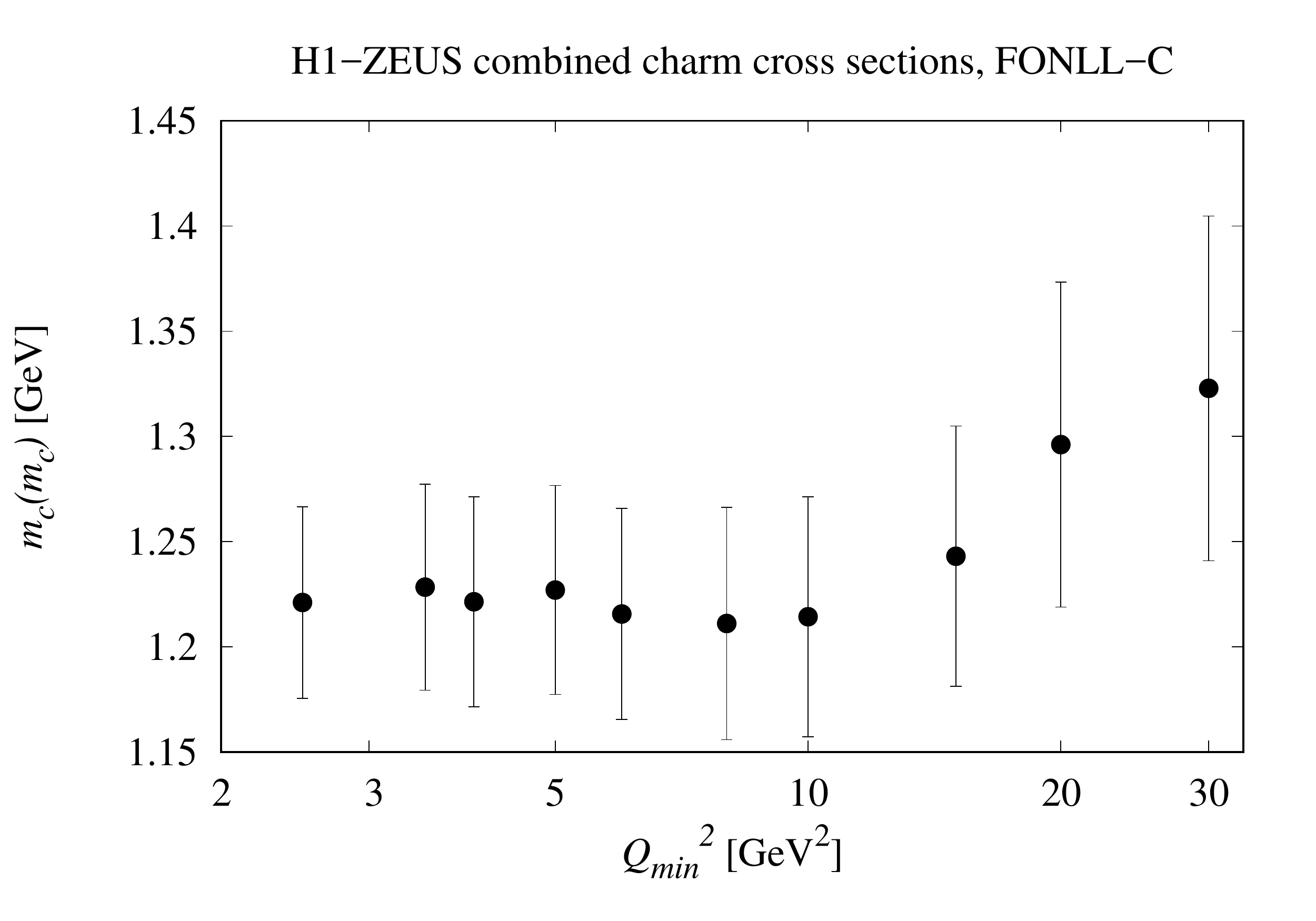}
  \end{center}
  \caption{\label{fig:q2min_f2ch} Left plot: dependence of
    $\Delta\chi^2/\Delta N_{\rm points}$ as a function of $Q_{\rm
      min}^2$ for the charm cross-section subset. Right plot:
    dependence of the best fit value of $m_c(m_c)$ with the associated
    experimental uncertainty as a function of $Q_{\rm min}^2$ for the
    charm cross-section subset in the combined fit. Both plots have
    been obtained using the FONLL-C scheme.}
\end{figure*}

In this context, it is interesting to look at the behaviour of the
partial $\chi^2$'s as a function of $Q_{\rm min}^2$ of the charm and
inclusive cross-section data separately to assess in a more specific
way which nominal value of $Q_{\rm min}^2$ is more convenient. Since
the meaning of ``degrees of freedom'' is unclear for a subset of the
full data set, in order to quantify the degree of improvement in the
partial $\chi^2$'s, we consider the following quantity:
\begin{equation}\label{eq:DeriveChi2}
\frac{\Delta\chi^2}{\Delta N_{\rm points}}(Q_{\rm min}^2) =
\frac{\chi^2(Q_{\rm min}^2) - \chi^2(Q_{\rm min}^2 = 2.5\mbox{ GeV}^2)}{N_{\rm points}(Q_{\rm min}^2) - N_{\rm points}(Q_{\rm min}^2 = 2.5\mbox{ GeV}^2)}\,,
\end{equation}
which provides an estimate of the improvement of the $\chi^2$ per data
point with respect to our lowest cut $Q_{\rm min}^2 = 2.5$ GeV$^2$. If
for a given value of $Q_{\rm min}^2$ this quantity is larger than one,
this means that that specific cut leads to an improvement of the
$\chi^2$ which is larger than the degrees of freedom subtracted by
excluding a given number of data points and thus the excluded data
points with respect of the reference cut (2.5 GeV$^2$) are poorly
described. On the contrary, if the quantity in
eq.~(\ref{eq:DeriveChi2}) is smaller than one, this means that the
excluded data points are better described than the fitted ones. In the
left panel of Fig.~\ref{fig:q2min_f2ch} we show the behaviour of the
contribution to the global $\Delta\chi^2/\Delta N_{\rm points}$
originating from the charm data points only. It is clear that any cut
between 3.5 and 5 GeV$^2$ improves drastically the partial $\chi^2$
while cuts above 5 GeV$^2$ either cause a much less significant
improvement or even lead to a deterioration. This provides a further
confirmation of the fact that our nominal cut (3.5 GeV$^2$) is a
sensible choice.

It is also interesting to look at the best fit values of $m_c(m_c)$
and the relative uncertainty preferred by a given subset as a function
of $Q_{\rm min}^2$ to quantify the sensitivity to $m_c(m_c)$ as more
and more data are excluded from the fit. This is plotted in the right
panel of Fig.~\ref{fig:q2min_f2ch} for the charm cross-section
data. It is clear that this particular subset of data tends to prefer
values of $m_c(m_c)$ around 1.23 GeV which is substantially lower than
the global value given in eq.~(\ref{eq:fonll}). The stability of the
central value of $m_c(m_c)$ for different values of $Q_{\rm min}^2$ is
remarkable and, as expected, the experimental uncertainty tends to
increase for larger value of $Q_{\rm min}^2$ indicating a loss of
sensitivity.

\begin{figure*}
  \begin{center}
    \includegraphics[angle=270,width=0.49\textwidth]{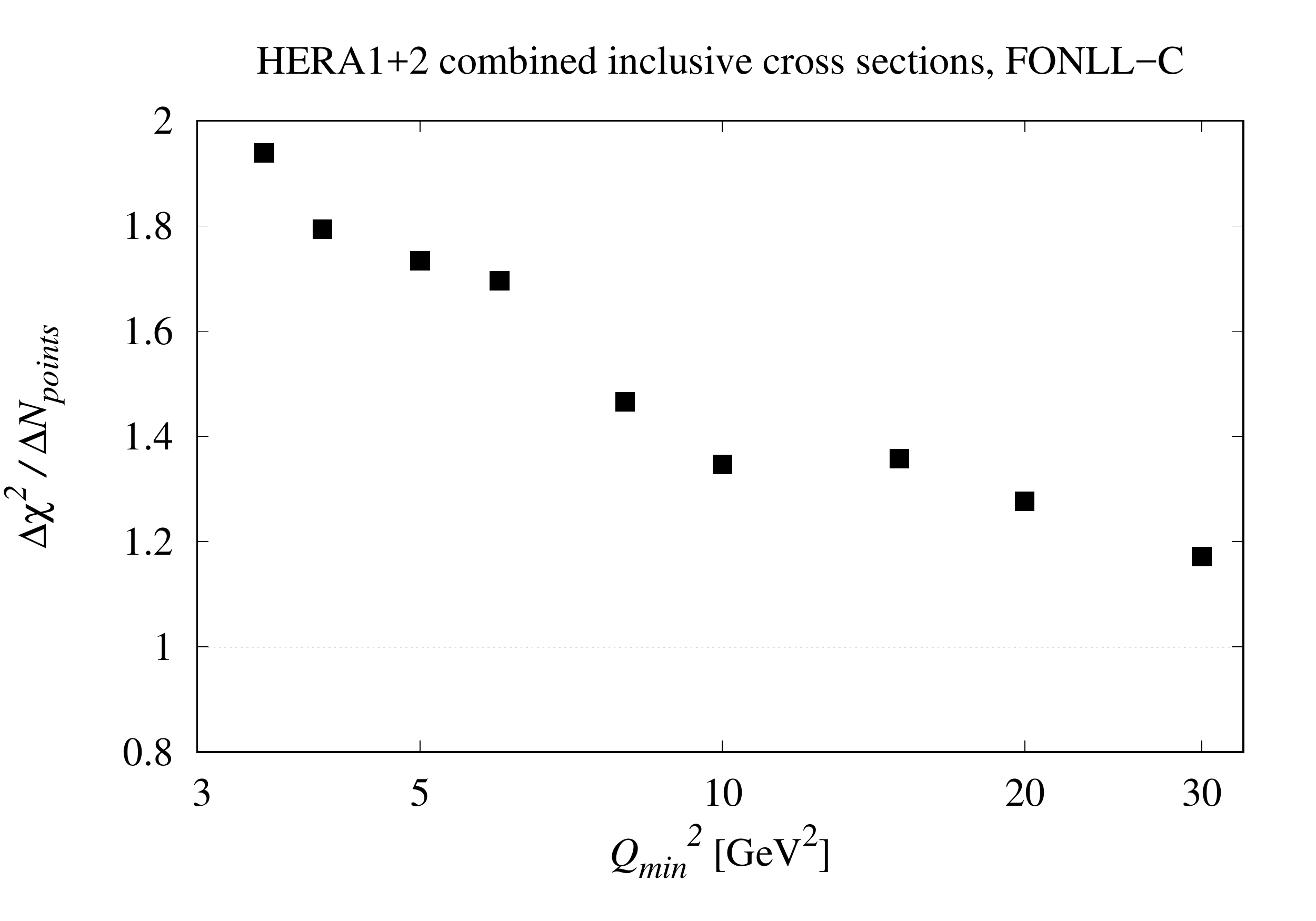}
    \includegraphics[angle=270,width=0.49\textwidth]{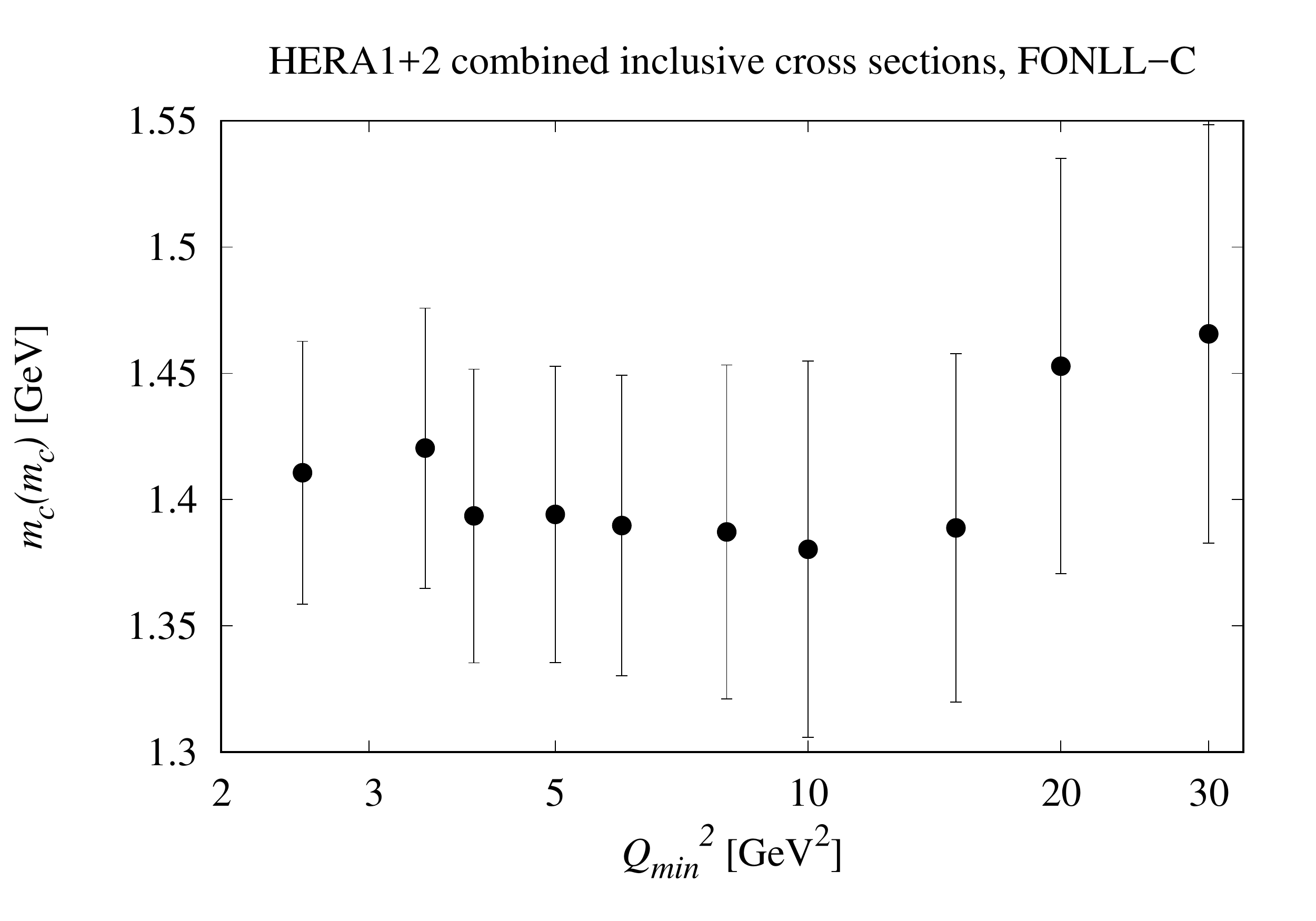}
  \end{center}
  \caption{\label{fig:q2min_incl} Same as Fig.~\ref{fig:q2min_f2ch}
    for the inclusive cross-section subset in the combined fit.}
\end{figure*}

Finally, we have done the same exercise for the HERA1+2 inclusive
cross-section data and in Fig.~\ref{fig:q2min_incl} we present the
relative plots. In the left panel we observe that the $\chi^2$ of this
subset improves essentially monotonically as $Q_{\rm min}^2$ increases
while from the right panel it is clear that the preferred value of
$m_c(m_c)$ of the inclusive cross sections is substantially larger
than that preferred by the charm cross sections with, again,
uncertainties than become broader for larger values of
$Q_{\rm min}^2$.  It is finally clear that our best value for
$m_c(m_c)$ quoted in eq.~(\ref{eq:fonll}) is a compromise between the
lower value preferred by the exclusive charm data and the larger value
preferred by the inclusive data.

\subsection{Discussion on the sensitivity to $m_c(m_c)$ of the
  inclusive data}\label{sec:IncDataSens}

\begin{figure*}
  \begin{center}
    \includegraphics[angle=270,width=0.7\textwidth]{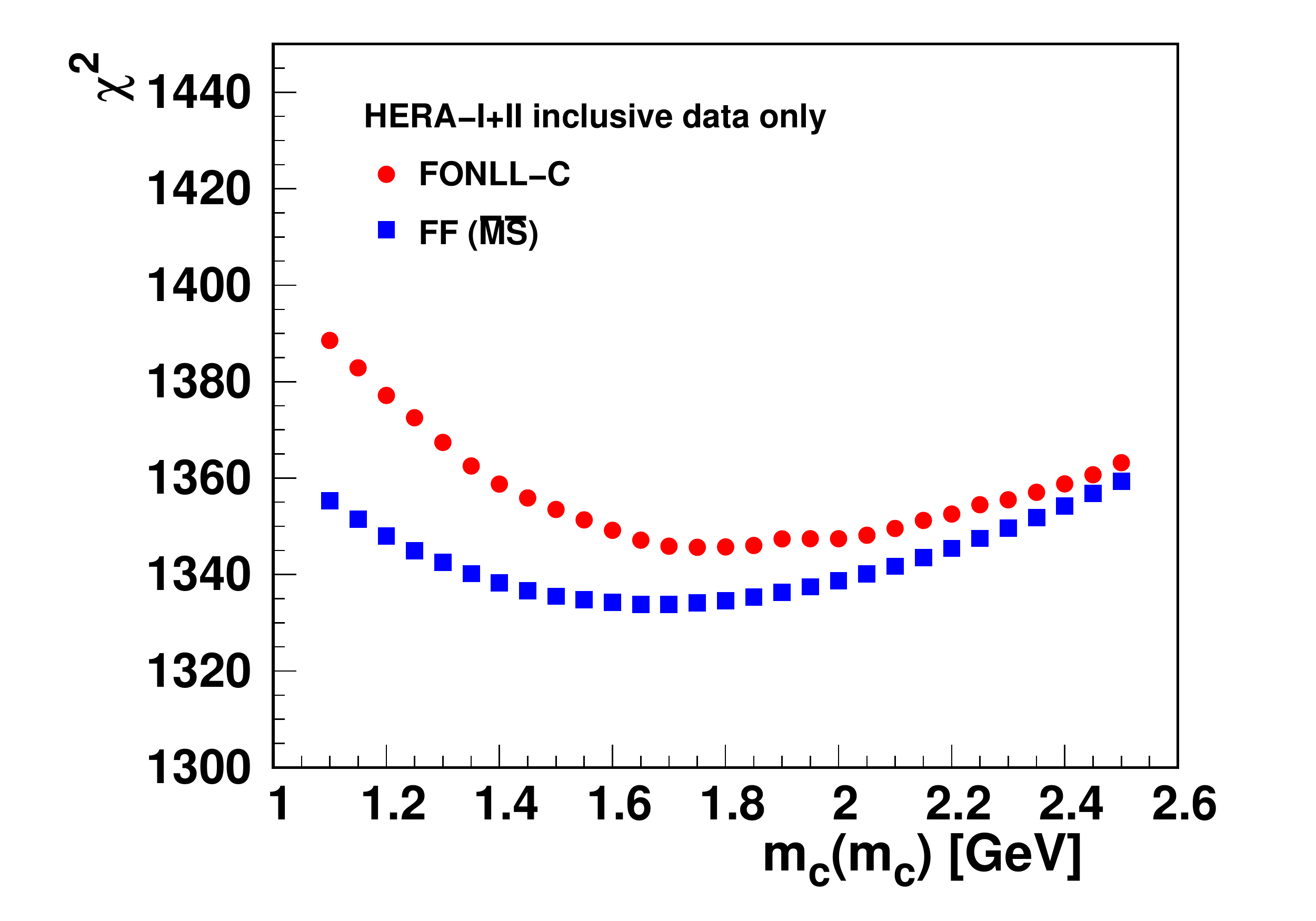}
  \end{center}
  \caption{$\chi^2$ vs. $m_c(m_c)$ profile for the fits to the HERA1+2
    inclusive cross sections only. The red circles indicate the
    FONLL-C scheme while the blue squares the FFN scheme at
    NLO.\label{fig:chi2inclusive}}
\end{figure*}

It is clear from the right panels of Figs.~\ref{fig:q2min_f2ch}
and~\ref{fig:q2min_incl} that the exclusive charm and inclusive data
subsets prefer somewhat different values of $m_c(m_c)$. However, the
values shown in these figures are clearly correlated because they were
obtained in a simultaneous fit to all data. In order to investigate a
possible tension, we have performed a fit to the inclusive data only
using both the FONLL-C and FFN schemes. The $\chi^2$ profiles are
shown in Fig.~\ref{fig:chi2inclusive}. In contrast to
Figs.~\ref{fig:fonllC} and~\ref{fig:ffABM}, in both schemes the scan
in $m_c(m_c)$ of the fits to inclusive data only yielded a shallow
$\chi^2$ dependences with a minimum around $1.7$ GeV. This
demonstrates that the inclusive data alone cannot constrain $m_c(m_c)$
reasonably well, but also why this data exerts an upwards pull on the
$m_c(m_c)$ value in the combined fit. Furthermore, since
Figs.~\ref{fig:q2min}, \ref{fig:q2min_f2ch}, and~\ref{fig:q2min_incl}
in Sect.~\ref{sec:Q2minDependence} present an overall remarkable
stability of the central value of $m_c(m_c)$ for different values of
$Q^2_{\rm min}$, the observed feature cannot be attributed to the low
$Q^2$ part of the inclusive data.

\section{Conclusions}\label{sec:summary}

In this work we have presented a new determination of the
$\overline{\rm MS}$ charm quark mass $m_c(m_c)$ obtained by fitting
HERA charm and inclusive DIS data. In particular, we included in our
fits the combined H1 and ZEUS charm production cross-section
measurements~\cite{Abramowicz:1900rp} and the final combination of
HERA1+2 H1 and ZEUS inclusive DIS cross-section
data~\cite{Abramowicz:2015mha}, the latter being used in this work for
the first time for the extraction of the charm mass.
Our determination is based on the FONLL general-mass
variable-flavor-number scheme, and has required the generalization of
the FONLL structure functions, originally constructed in the pole-mass
scheme, in terms of $\overline{\rm MS}$ heavy quark masses.

A detailed estimate of the various sources of uncertainty that affect
our determination of $m_c(m_c)$ has been performed.  In particular, we
estimated the uncertainties due to the choice of the PDF
parametrization, the model parameters used as input for the
theoretical computations, and the missing higher-order corrections. We
found that those sources of uncertainty are smaller than the
experimental uncertainty, resulting in a competitive determination of
the charm mass.

We complemented the FONLL extraction of the charm mass with an
analogous determination based on the fixed-flavour number scheme at
next-to-leading order, finding a good agreement between the two. In
addition, we compared our results with previous determinations also
based on fits to DIS data and with the PDG world average finding again
a generally good agreement. We find that the values extracted in this
work, although compatible within uncertainties, tend to be slightly
higher than previous determinations from HERA data. This feature seems
to be associated to the final HERA1+2 combined inclusive dataset,
which tends to prefer larger values of $m_c(m_c)$ as compared to the
charm structure function data, and thus increases the best-fit value.

In the future, it would be interesting to repeat the FONLL
determination in the context of a global PDF analysis, since, in
addition to the inclusive and charm HERA data, other experiments are
expected to have some sensitivity to the value of the $\overline{\rm
  MS}$ charm mass.
In addition, the use of a wider dataset might lead to a reduction of
the experimental uncertainties of the $m_c(m_c)$ determination.
Moreover, our analysis is based on the standard assumption that the
charm PDF is dynamically generated by collinear splitting from gluons
and light quarks.
In this respect, it would be useful to redo the determination of
$m_c(m_c)$ in the presence of a possible non-perturbative charm PDF,
for which the generalized FONLL structure functions accounting for a
fitted heavy quark PDF are available~\cite{Ball:2015tna}.

\acknowledgments We would like to thank Fred Olness for the careful
reading of the manuscript and for providing helpful comments and
suggestions. J.~R. is supported by an STFC Rutherford Fellowship and
Grant ST/K005227/1 and ST/M003787/1.  V.~B. and J.~R. are supported by
the European Research Council Starting Grant ``PDF4BSM''.  We are
grateful to the DESY IT department for their support of the {\tt
  xFitter} developers. 


\bibliography{HERAfitterFONLL.bib}

\end{document}